\documentclass{emulateapj}
\usepackage{natbib}
\usepackage{ulem}

\shorttitle{NGVS. V. Modelling the dynamics of M87}
\shortauthors{Zhu et al.}

\begin{document}
\title{The Next Generation Virgo Cluster Survey. V. Modelling the dynamics of M87 with the made-to-measure method}
  
  \author{Ling~Zhu\altaffilmark{1,2,16}}
  \author{R. J.~Long\altaffilmark{1,3}}
  \author{Shude~Mao\altaffilmark{1,3}}
   \author{Eric~W. Peng\altaffilmark{4,5}}
   \author{Chengze Liu\altaffilmark{6,15}}
   \author{Nelson Caldwell\altaffilmark{7}}
   \author{Biao Li \altaffilmark{4}}
   \author{John P. Blakeslee\altaffilmark{8}} 
   \author{Patrick C\^ot\'e\altaffilmark{8}}
   \author{Jean-Charles Cuillandre\altaffilmark{9}}
   \author{Patrick Durrell\altaffilmark{10}}
   \author{Eric Emsellem \altaffilmark{11,12}}
   \author{Laura Ferrarese \altaffilmark{8}}
   \author{Stephen Gwyn \altaffilmark{8}}
   \author{Andr\'es Jord\'an \altaffilmark{13}}
   \author{Ariane Lan\c{c}on\altaffilmark{14}}
   \author{Simona Mei\altaffilmark{17,18}}
   \author{Roberto Munoz\altaffilmark{13}}
   \author{Thomas Puzia\altaffilmark{13}}

   \altaffiltext{1}{National Astronomical Observatories, Chinese Academy of Sciences, A20 Datun Rd, Chaoyang District, Beijing 100012, China}
    \altaffiltext{2}{Max-Planck-Institut f�r Astronomie, K�nigstuhl 17, D-69117 Heidelberg, Germany,  \email{lzhu@mpia-hd.mpg.de}}
   \altaffiltext{3}{Jodrell Bank Centre for Astrophysics, Alan Turing Building, The University of Manchester, Manchester M13 9PL, UK}
   \altaffiltext{4}{ Department of Astronomy, Peking University, Beijing 100871, China}
\altaffiltext{5}{ Kavli Institute for Astronomy and Astrophysics, Peking University, Beijing 100871, China}
\altaffiltext{6}{ Department of Physics and Astronomy, Shanghai Jiao Tong University, 800 Dongchuan Road, Shanghai 200240, China}
\altaffiltext{7}{Harvard-Smithsonian Center for Astrophysics, Cambridge, MA 02138, USA}
\altaffiltext{8}{ National Research Council of Canada, Victoria, BC V9E 2E7, Canada}
\altaffiltext{9}{ Canada-France-Hawaii Telescope Corporation, Kamuela, HI 96743, USA}
\altaffiltext{10}{ Department of Physics and Astronomy, Youngstown State University, One University Plaza, Youngstown, OH 44555, USA}
\altaffiltext{11}{ Universit\'e de Lyon 1, CRAL, Observatoire de Lyon, 9 av. Charles Andr\'e, F-69230 Saint-Genis Laval; CNRS, UMR 5574; ENS de Lyon, France}
\altaffiltext{12}{ European Southern Observatory, Karl-Schwarzchild-Str. 2, D-85748 Garching, Germany}
\altaffiltext{13}{ Instituto de Astrof\'{i}sica, Facultad de F\'{i}sica, Pontificia Universidad Cat\'{o}lica de Chile, Av. Vicu\~{n}a Mackenna 4860, 7820436 Macul, Santiago, Chile}
\altaffiltext{14}{ Observatoire astronomique de Strasbourg, Universit\'e de Strasbourg, CNRS, UMR 7550, 11 rue de l�Universite, F-67000 Strasbourg, France}
\altaffiltext{15}{ INPAC, Department of Physics and Astronomy and Shanghai Key Lab for Particle Physics and Cosmology, Shanghai Jiao Tong University, Shanghai, 200240, China}
\altaffiltext{16}{ Department of Physics and Tsinghua Centre for Astrophysics,Tsinghua University, Beijing 100084, China}   
\altaffiltext{17}{ GEPI, Observatoire de Paris, 77 av. Denfert Rochereau, F-75014 Paris, France}
\altaffiltext{18}{ Universit\'{e} Paris Denis Diderot, F-75205, Paris Cedex 13, France}

\begin{abstract}
We study the dynamics of the giant elliptical galaxy M87 from the central to the outermost regions with the made-to-measure (M2M) method. We use a new catalogue of 922 globular cluster line-of-sight velocities extending to a projected radius of 180 kpc (equivalent to 25 M87 effective radii), and SAURON integral field unit data within the central 2.4 kpc. 
263 globular clusters, mainly located beyond 40 kpc, are newly observed by the Next Generation Virgo Survey (NGVS).
For the M2M modelling, the gravitational potential is taken as a combination of a luminous matter potential with a constant stellar mass-to-light ratio and a dark matter potential modelled as a logarithmic potential. Our best fit dynamical model returns a stellar mass-to-light ratio in the $I$ band of $M/L_I=6.0\pm0.3$ $M_{\odot}/L_{\odot}$ with a dark matter potential scale velocity of $591\pm 50$ km s$^{-1}$ and scale radius of $42 \pm 10$ kpc. We determine the total mass of M87 within 180 kpc to be $(1.5\pm 0.2)\times 10^{13} M_{\odot}$. The mass within 40 kpc is smaller than previous estimates determined using globular cluster kinematics that did not extend beyond $\sim 45$ kpc. With our new globular cluster velocities  at much larger radii, we see that globular clusters around 40 kpc show an anomalously large velocity dispersion which affected previous results.
The mass we derive is in good agreement with that inferred from \emph{ROSAT} X-ray observation out to 180 kpc. Within 30 kpc our mass is also consistent with that inferred from \emph{Chandra} and \emph{XMM-Newton} X-ray observations, while within 120 kpc it is about $20\%$ smaller.
The model velocity dispersion anisotropy $\beta$ parameter for the globular clusters in M87 is small, varying from $-0.2$ at the centre to 0.2 at $\sim40$ kpc, and gradually decreasing to zero at $\sim120$ kpc. 
\end{abstract}

\keywords{
  Galaxy: kinematics and dynamics ---
  Galaxy: M87---
  Galaxy: globular cluster---
  Method: made-to-measure}

\section{Introduction}
\label{sec:intro}
Elliptical galaxies lack the dynamical simplicity of cold stellar disks in spiral galaxies,  posing well-known challenges in determining their intrinsic properties (mass distribution, shape, orbit structure). 
Their dark matter distributions are even harder to determine since they only start to dominate at several effective radii where the galaxies become too faint for efficient measurement of their stellar dynamics in integrated light.  
Line-of-sight (LOS) velocities of discrete Globular Clusters (GCs) and Planetary Nebulae (PNs) can be measured in more extended regions of galaxies, even to the edge of their dark matter halos, and provide excellent tracers of the outer dark matter distributions.

M87, the second brightest giant elliptical galaxy in Virgo, is located at the dynamical centre of the cluster. The formation history of such galaxies is of keen interest in structure formation studies. Their formation through dry mergers in a cluster environment has been extensively discussed (e.g. \citealt{del2007, bern2007, Liu2009, Liu2012}). A related question is whether the dark matter distribution at the centre of Virgo follows an equilibrium smooth profile or exhibits substructure. The early type Virgo galaxies' velocity dispersion profiles indicate that the dark matter inside 2 Mpc is smooth and dynamically relaxed \citep{McLaughlin1999}. However, the PN LOS velocity dispersion decreases to $78\pm25$ km s$^{-1}$ at 144 kpc, indicating that the stellar halo of M87 may be truncated at this radius \citep{Doherty2009}. This result, however, is based on only 12 PN LOS velocities, and thus may be influenced by incomplete PN sampling effects. In this work, as a first step, we treat M87 as a relaxed, equilibrium system.

M87 has the richest GC system in the local supercluster with a total of 15000 GCs (\citealt{McLaughlin1994}; \citealt{Peng2008}).
The database of M87 GCs with LOS velocities has been growing rapidly in recent years (Durrell et al. 2014 in preparation; see \S5 for more complete references and discussions). Their kinematics, together with central stellar kinematics, have been used to derive M87's dark matter distribution, mostly using the
\cite{Schwarzschild1979} technique (e.g. \citealt{Murphy2011}, hereafter M2011; \citealt{Gebhardt2009}, hereafter G2009). The dark matter halo and stellar parameters so determined are often different. This is largely a reflection of the fact that the models are degenerate in their mass determination, lacking separate constraints on the dark and luminous matter components. In \S5, we compare these earlier results with ours in greater detail\footnote{Notice that
different values of the distance to M87 are used in the literature.  In the preceding paragraphs, we have converted all the radii from \arcsec\, to kpc with the distance we have adopted, $d=16.5$ Mpc.}.

M87 also has diffuse X-ray emission from hot gas. There are X-ray observations of the galaxy extending to large radii. Although the hot gas may not be in complete hydrostatic equilibrium, the \emph{ROSAT} X-ray observations of M87 has been used to derive the mass profile of M87 out to a radius of $\sim 270$ kpc \citep{Nulsen1995}, while {\it Chandra} and {\it XMM-Newton} X-ray observations have been used to derive the mass profile of M87 out to a radius of $\sim 120$ kpc (\citealt{Churazov2010}; \citealt{Das2010}). The mass derived using X-ray observations is less than that from GCs. 

The GCs previously used to derive M87's mass extended to a maximum radius of 45 kpc, with a rising LOS velocity dispersion. These GCs show medium rotation along the minor axis with some possible kinematical substructures around $40$ kpc where the stellar dynamics data are not accessible. GC dynamics thus provide a crucial input in this region.
Furthermore, 
GCs extending to the extreme outer regions of M87 are crucial for studying the dark matter halo of M87 out to the approximate edge of its gravitational potential boundary. For example, \cite{Strader2011} obtained new, precise radial velocities for 451 M87 GCs with projected radii from $\sim5$ to 185 kpc. They found much lower values than earlier papers for the velocity dispersion and rotation of the GCs within the region the previous data covers  (see \S5), and also differed from previous works in seeing no evidence for a transition from the inner halo to a potential dominated by the Virgo Cluster, nor for a truncation of the stellar halo in the outer region. 

In this paper, we construct made-to-measure (M2M) models (\citealt{Hunt2013}; \citealt{Long&Mao2010}; \citealt{deLorenzi2007}; \citealt{Syer1996}) to investigate the dynamics of the elliptical galaxy M87 using kinematic measurements of its GCs. In so doing, we take an early intercept of data from the NGVS programme.  Other papers in the NGVS series, related to the topics considered here, include those on the distribution of globular clusters in Virgo (Durrell et al. 2014), the properties of star clusters, ultra-compact dwarfs (UCDs) and galaxies in the cluster core (Peng et al. 2014; Liu et al. 2014), interactions within possible in-falling galaxies (Paudel et al. 2013), and optical-IR source classification (Mu$\widetilde{n}$oz et al. 2014).

The paper is arranged as follows. 
In \S 2, we first describe the observational data for M87 which will be used to constrain the M2M models, including discrete GC LOS velocities out to 180 kpc and SAURON IFU data within 2.4 kpc.
In \S 3, we introduce our construction of M2M models for elliptical galaxies, particularly when using discrete data as model constraints. 
In \S 4, we focus on M87 describing the modelling steps taken and our results for M87. We discuss our results in \S 5, and draw conclusions in \S 6 outlining some areas for improvement in future investigations.

\section{Observational data}
In this section, we describe the M87 observational data that we will use for modelling. 
We adopt a distance to M87 of 16.5 Mpc (\citealt{Blakeslee2009}; \citealt{Mei2007}), and an $I$-band magnitude $M_{\mathrm I} = 7.23$ \citep{Cappellari2006}. At the distance of M87, 1 arcs corresponds to a physical scale of 80 pc.

\subsection{Photometric and number density data}
\label{sec:phdata}
\begin{figure}
\centering\includegraphics[width=8.2cm, height=7.7cm]{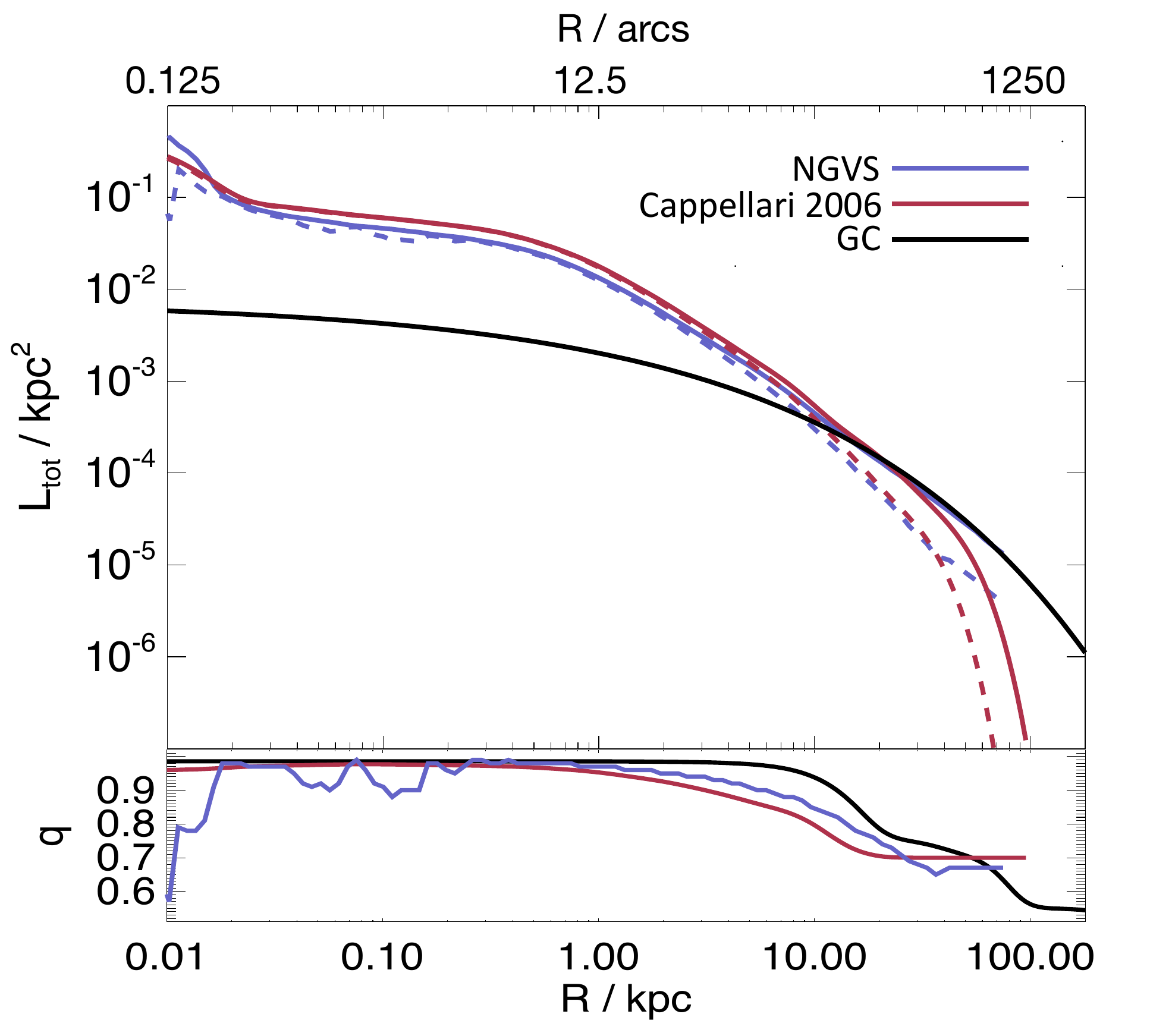}
\caption{{\bf Top}: the normalized surface density profiles. The black line is the GC surface number density profile along projected radius from  \protect\cite{Peng2008}, the red solid and dashed line are the stellar surface brightness profile along major and minor axis constructed from the M87 MGE in \protect\cite{Cappellari2006}, the purple solid and dashed lines are the  g band surface brightness profile along major and minor axis from NGVS (Ferrarese et al., in preparation). The total luminosity/number for each profile within 80 kpc have been normalized to be unity. {\bf Bottom}: the minor to major axis ratio $q$ of the GC surface number density,  \cite{Cappellari2006} and NGVS stellar surface brightness, respectively.}
\label{fig:SB}
\end{figure}

We assume a spherical spatial distribution of the M87 GCs. The surface number density profile can be well fitted by a Sersic function:
\begin{equation}
\label{eqn:sersic}
\log I \sim -b_n\Bigr [\Bigr (\frac{R}{R_0}\Bigr )^{\frac{1}{n}}-1.0 \Bigr] + \mathrm {constant},
\end{equation}
where $R$ is the projected radius, $I$ is the surface brightness, $R_0$ is the scale radius and $n$ is the Sersic index. The current best fit to the GC surface number density profile is $b_n = 2.21$, $R_0 = 510.24''\pm 45.31$, $n = 2.71\pm 0.17$ \citep{Peng2008}, which was derived based on all the photometrically detected GCs ($\sim 15000$ in total) from the ACS Virgo Cluster Survey (\citealt{Cote2004}; \citealt{Ferrarese2006}), and not just those GCs with kinematic measurements.  The black line in Fig.~\ref{fig:SB} shows the GC surface number density profile along the projected radius.  

The stellar surface density profile is different from that of the GCs in M87, with the stars being more concentrated in the central 10 kpc than the GCs.  We assume an axisymmetric spatial distribution of the stars, and use a 2D multi-Gaussian expansion fit to the photometric image of M87. The red line in Fig.~\ref{fig:SB} is the surface brightness along the major axis constructed from the MGE fit performed in \cite{Cappellari2006}, based on the {\it Hubble Space Telescope (HST)/WFPC2} and ground-based MDM photometry in the $I$ band. The MDM observations have a field of view of $17.1\times 17.1$ arcmin$^2$,
which covers the region $R<40$ kpc for M87.  It is well known that the isophotes of M87 become progressively flattened at large radii (e.g., \citealt{Carter1978}; \citealt{Caon1990}; \citealt{Mihos2005}; \citealt{Kormendy2009}; \citealt{Janowiecki2010}).  The MGE of the surface brightness we are using is consistent with this flattening.

NGVS imaging covers a much wider field of view (the entire Virgo Cluster), and allows improvements in both the GC surface number density and stellar surface brightness profile. However, because the NGVS profile of M87 (Ferrarese et al., in preparation) was unavailable when this work was started, we chose to use the \cite{Cappellari2006} surface brightness profile and the spherical GC surface number density profile \citep{Peng2008} in our models. The new profiles based on NGVS are shown in Fig.~\ref{fig:SB}.  

There is evidence that the GC number density is not really spherical ( \citealt{Strader2011}; \citealt{Forte2012};  \citealt{McLaughlin1994}; Durrell et al. 2014, ApJ, submitted).  NGVS imaging detected more GCs than before so a 2D number density profile become possible. The black line in the bottom panel of Fig.~\ref{fig:SB} shows the flattening of the GC distribution derived from NGVS imaging.  

The potential impact on our models of the new surface brightness and number density profiles and the ellipticity of the GC number density are discussed later in \S 5.  In future investigations, we plan to revisit the dynamics of M87 using both NGVS optical-IR surface photometry and an expanded sample of globular cluster velocities.

The fact that the M87 GCs and stars follow different surface density profiles indicates that they also follow different distribution functions. Even though their dynamics follow the same potential, we construct two individual sets of models to model the GCs and stellar dynamics separately, as in previous investigations (e.g. \citealt{Murphy2011}). The GC surface number density profile from \protect\cite{Peng2008} (as in Fig.~\ref{fig:SB}) is used as an observational constraint for the M2M GC models. 
The MGE surface brightness from \cite{Cappellari2006} is used as an observational constraint for the M2M stellar models. The stellar models do not extend to large radii because they are limited by SAURON kinematics which only extend to $R\sim2.4$ kpc. 

\subsection{Kinematical data}

\subsubsection{GC kinematics}
\label{sec:GCki}
\cite{Strader2011} merged previous observations of M87 GCs with their own observations to produce a catalogue of 737 GCs with LOS velocity measurements. To this we have added 263 new GCs with velocity measurements (299 in total less 36 duplicates already in the catalogue),
most of which are at large galactocentric radii (beyond 40 kpc). The new GCs were selected from imaging in the Next Generation Virgo Survey (NGVS) \citep{Ferrarese2012}, which is a Large Programme on the {\it Canada France Hawaii Telescope} (CFHT) to produce deep optical, multiband photometry of the entire Virgo Cluster of galaxies out to the virial radius. The contiguous coverage of NGVS makes it ideally suited to finding and studying GCs in the outer halos of galaxies and intracluster regions.

\begin{figure}
\centering\includegraphics[width=8cm, height=6cm]{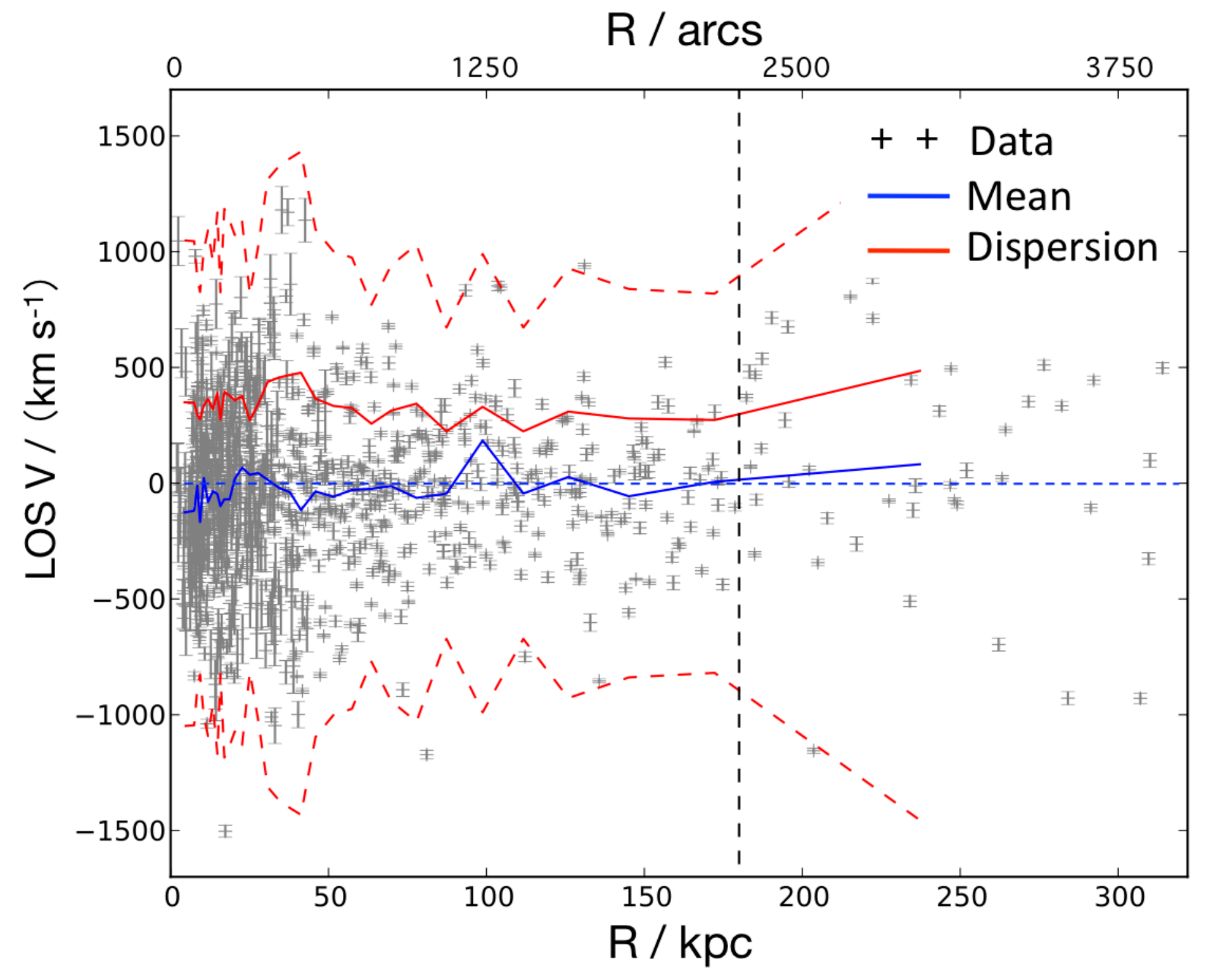}
\caption{The GC LOS velocity distribution along the projected radius $R$. The blue solid line indicates the mean GC velocity profile, the red solid line shows the velocity dispersion, and the red dashed lines indicate the $3\sigma$ boundary of the velocity distribution. The blue dashed line represents $v=0$. The vertical dashed line at $R=180$ kpc indicates the radial limit of the data used in the M2M models.}
\label{fig:GC1}
\end{figure}

Using NGVS imaging, we used colour and morphological criteria to select candidate GCs around M87 with $g<22.5$ mag, and  obtained spectroscopy for them using Hectospec \citep{Fabricant2005}, the multifibre spectrograph on the MMT on Mt. Hopkins. We reduced the spectra using the standard SAO pipeline, and obtained radial velocities by cross-correlating the spectra with stellar and GC templates. The details of the observations will be presented in a future paper (Peng et al. 2014). We combined these data with those in the literature to produce a sample of 966 GCs inside 320 kpc (922 GCs within 180 kpc). Although the final sample of GCs is from heterogeneous sources, cross-checks of overlapping sources indicate that they are consistent within the error bars. Duplicate observations were combined as appropriate to produce a velocity estimate for each object. NGVS has contributed significantly to the number of GCs at larger radii.  
For example, the number of GCs with kinematic measurements outside of 40 kpc is now in excess of 440, an increase of about $50\%$ due to NGVS.

\begin{figure}
\centering\includegraphics[width=\hsize]{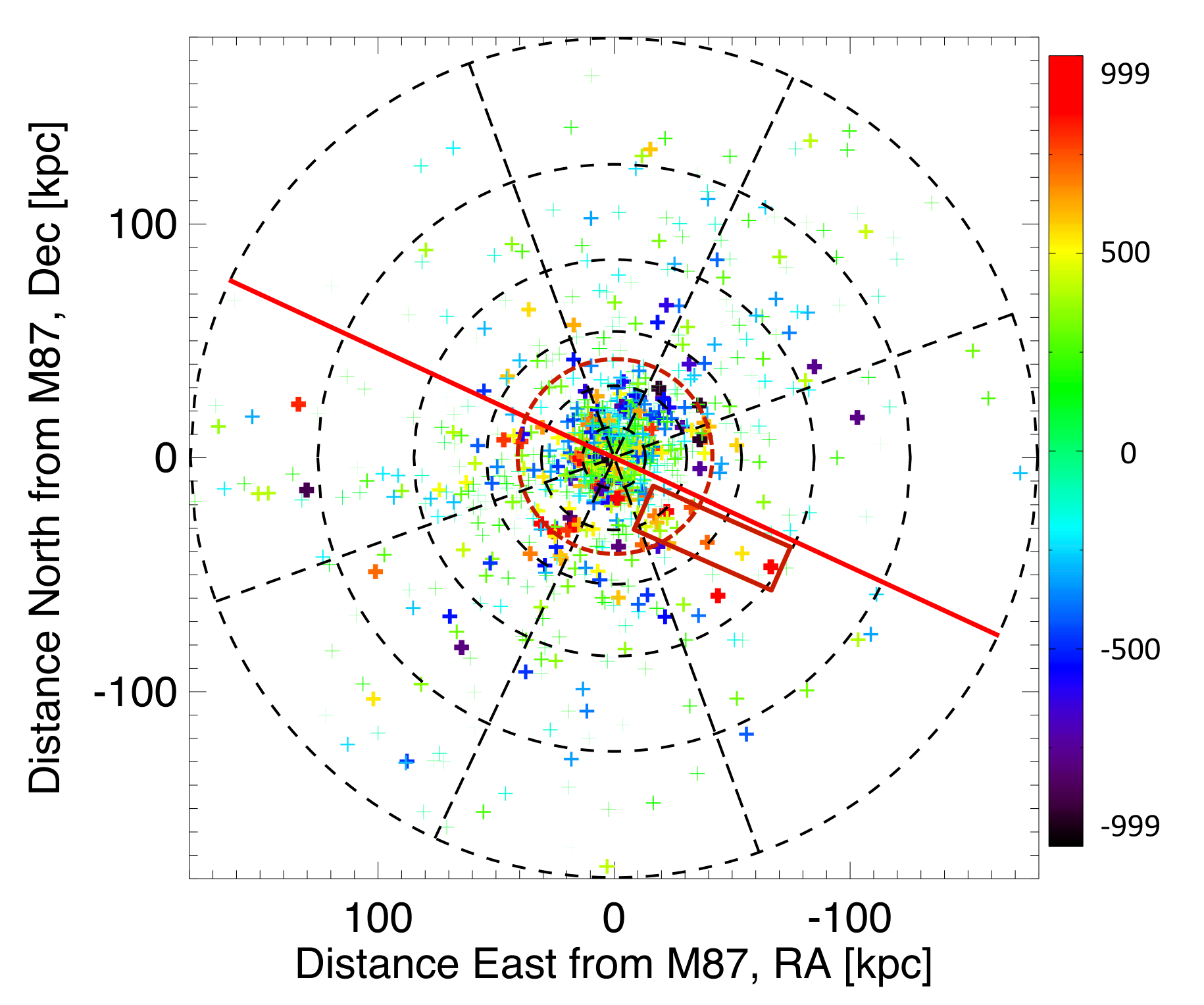}
\caption{The 2D LOS velocity map for the 922 GCs within 180 kpc (before the number of GCs is reduced to 896 by $3\;\sigma$ clipping and substructure removal). RA and Dec are in units of kpc, and the red line indicates the orientation of the minor axis of M87. Each plus symbol represents a GC data point and the colours indicate the values of the velocities. The red dashed circle has radius $R = 40.8$ kpc and the red rectangle indicates a possible substructure. The colour bar gives the velocity range in units of km s$^{-1}$. The figure also illustrates the particle binning scheme for discrete data models. In this illustration, there are 6 radial and 8 angular divisions with some discrete data points located in each bin. For our M2M models, we use 16 radial and 16 angular divisions.}
\label{fig:GCk}
\end{figure}
Fig.~\ref{fig:GC1} shows the velocity distribution of the GCs along the projected radius $R$ with M87's systemic velocity of 1274 km s$^{-1}$ \citep{Emsellem2004} having been subtracted. 
Each data point represents a GC, the solid blue line is the mean velocity profile, the solid red line is the velocity dispersion. The mean velocity and velocity dispersion profiles are calculated using equal count binning and are not smoothed.  
The GCs inside 180 kpc have a mean velocity fluctuating around zero (see Fig.~\ref{fig:GC1}),
and the spatial distribution of the GCs is nearly uniform in radius and azimuth (see Fig.~\ref{fig:GCk}). 
Outside 180 kpc, the mean velocity deviates slightly from zero and the velocity dispersion becomes larger. This perhaps indicates that GCs outside 180 kpc may not be gravitationally bound by the potential of M87, i.e. they may be affected by the potential of nearby galaxies. However, for the whole sample, deviations of the mean GC velocity from zero are not significant statistically. This situation is similar to that seen in the PN velocity distribution \citep{Doherty2009}. We chose therefore to use only the GCs inside 180 kpc to trace the potential of M87. 
For the 922 GCs inside 180 kpc, we perform $3\sigma$ velocity clipping to tidy the data before using them in the M2M models. As a result, 6 GCs with velocities outside of $3\sigma$ from the mean velocity distribution are excluded (see Fig.~\ref{fig:GC1}).  

Fig.~\ref{fig:GCk} shows the positions of the GCs coloured according to their LOS velocity values. 
There is a large velocity dispersion and an obvious rotation around the minor axis (indicated by the red line in Fig.  \ref{fig:GCk}) at $R \approx 40$ kpc in agreement with previous results (\citealt{Strader2011}; \citealt{Cote2001}; \citealt{Cohen2000}). Some of the highest velocities in this region have been identified as measurement errors \citep{Strader2011}, and have been excluded from our sample. The large observed velocity dispersion, however, is not due only to this effect, as we discuss below.

The extended stellar envelope of M87 may have been built up by continual mergers, and as a result, the GCs may not have reached virial equilibrium \citep{RomanowskyM872012}. 
Around the region $R \approx 40$ kpc, the GCs show strong fluctuations in kinematics and relatively large rotation about the minor axis.  
From the velocity colouring in Fig.~\ref{fig:GCk}, there are more substructures around this region than any other region  and this may be part of the reason for the high velocity dispersion. Some of the substructures are quite obvious, for example the region marked by the red rectangle in Fig.~\ref{fig:GCk}, within which all 20 GCs have positive velocities. In this case, we choose to exclude these 20 GCs from our modelling. After exclusion, the large velocity dispersion around 40 kpc still exists.  

\begin{figure}
\centering\includegraphics[width=\hsize]{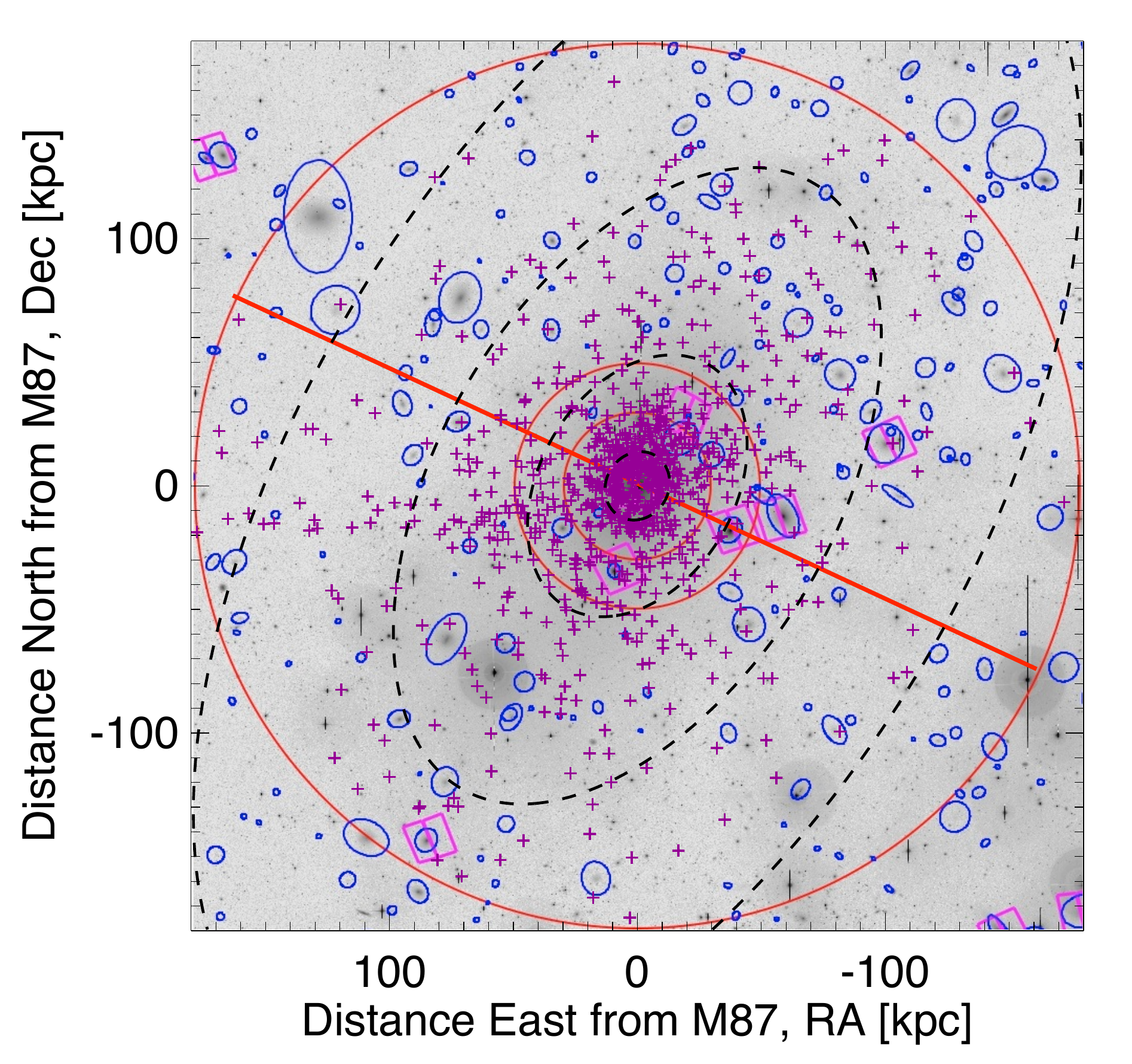}
\caption{Image of the Virgo centre over-plotted with the positions of GCs. Each magenta plus sign represents a GC with velocity measurement. The dashed ellipticals indicate the contours of the GC surface number density. The central Virgo galaxies in this region have been marked out with blue ellipses. The three red circles indicate $R = 30, 50,180$ kpc, and the red line indicates the orientation of the optical minor axis of M87.}
\label{fig:Vcc}
\end{figure}
 
As shown in Fig.~\ref{fig:Vcc}, there are a number of Virgo Cluster Catalogue (VCC) galaxies in the GC region,
but we can not match the offending GCs with any of the nearby (in projection) VCC galaxies.
The velocity differences between the VCC galaxies and the GCs in that region are off by many hundreds of km/s, if not more. 
It is not clear however that the kinematic fluctuations represent a real phenomenon.
A companion paper (Peng et al. 2014) will include a comprehensive examination of possible substructures and/or kinematic features in our sample. We proceed by reducing the impact of these data on our modelling by increasing the individual velocity errors of the 275 GCs in the region $20.4 < R < 57$ kpc by a factor of 5.0.

After the above reduction, we obtain 896 GCs inside 180 kpc. For the axisymmetric models we are going to build, we  symmetrize the data. 
The minor axis of M87 is indicated as the red line in Fig.~\ref{fig:GCk}.
We put the data in a coordinate system such that the Y axis is along the minor axis while the X axis is along the major axis.
Every GC with position coordinates (X, Y) on the sky and LOS velocity $V$ is reflected with respect to the centre of the galaxy to generate a new GC with  (-X, -Y, -V) and reflected with respect to the Y axis to generate a new GC with (-X, Y, -V).  This reflection will help to reduce further any GC subpopulation biases which might still be present. Symmetrisation is a common procedure used with many different modelling methods (e.g. \citealt{Cappellari2006}; \citealt{DeLorenzi2008}).

After the above reductions and symmetrisation, there are 3584 GC measurement points inside 180 kpc which will be used in the M2M models as discrete kinematical constraints.

\subsubsection{Stellar kinematics}
We take kinematic data for M87 from the SAURON data release \citep{Emsellem2004}. The data available are the LOS mean velocity $v$, velocity dispersion $\sigma$ and the $h_3$ and $h_4$ Gauss-Hermite coefficients, all taken from a truncated Gauss-Hermite expansion of the LOS distribution \citep{vanderMarel1993}. 
\cite{Long2012} modelled the SAURON data of M87 with the M2M method. We refer to \cite{Long2012} for the appropriate detail. We follow their data and modelling process but also include a dark matter potential as  will be described in \S~\ref{sec:dark}.

The SAURON observations were adaptively binned and processed in a centroidal Voronoi tessellation \citep{Cappellari2003}. M87 has 2112 such Voronoi bins.
Voronoi bins are not used for surface brightness. Instead we employ a $16\times 16$ polar grid with pseudo-logarithmic radial bins as described in \cite{Long&Mao2010}.

\section{Model construction}

\subsection{The made-to-measure method}
\label{sec:m2m}
The made-to-measure (M2M) method \citep{Syer1996} is particle based, rather than orbit based as in Schwarzschild's method (\citealt{Schwarzschild1979}; \citealt{Schwarzschild1993}). 
It is in other respects not dissimilar to the widely used Schwarzschild's method for dynamical modelling. 
For our purposes, the gravitational potential comes from a combination of luminous matter and dark matter and contains the parameters we want to determine. Weighted test particles are orbited within this potential and by modifying the particle weights at the same time, we aim to produce a weighted particle system in which the weights have individually converged to constant values, and the particle system is able to reproduce the measured observables of a real galaxy. By varying the parameters we are interested in and running multiple models, we are able to determine, in a maximum likelihood sense, the best fitting parameters. 

In recent years, 
the M2M method has been developed to model discrete data directly, without any binning prior to modelling (\citealt{Long&Mao2010}; \citealt{deLorenzi2007}; \citealt{DeLorenzi2008}; \citealt{Das2011}; \citealt{Hunt2013}). We use the M2M code developed initially in \cite{Long&Mao2010}, which is able to model both the IFU data and discrete data.


The key equation that leads to weight evolution over model time is
\begin{equation}
\label{eqn:Fe}
F = - \frac{1}{2}\chi_{LM} ^2 +  \frac{1}{\epsilon} \frac{d}{dt} S \mbox{} + \mu S - \frac{\lambda_{\rm sum}}{2}(\sum_i^Nw_i -1)^2, 
\end{equation}
where $\epsilon$, $\mu$ and $\lambda_{\rm sum}$ are positive parameters, $S=-\sum _i^N w_i \ln (\frac{w_i}{m_i} )$ is the entropy of the system, the $m_i$ are the prior weights and the $w_i$ are the particle weights and are considered to vary with time.  The $\chi_{LM} ^2$ term compares the observed data with equivalent values calculated from the weighted particles and is thus a function of the weights.

By maximising F with respect to the particle weights ($\partial F/\partial w_i =0$, $\forall  i$), we obtain the weight evolution equations in the form
\begin{equation}
\frac{d}{dt}w_i = -\epsilon w_i Q(\mathbf{w}).
\end{equation}
The overall speed of weight convergence is controlled by $\epsilon$, and $Q(\mathbf{w})$ is related to the observational constraints. We do not describe the details here (see \citealt{Long&Mao2010}). 

With several observational constraints in a model, we need to balance the effect of each constraint within the weight evolution equations. For $K$ multiple observables, we take $\chi^2_{LM}$ in the form
\begin{equation}
\label{eqn:chi2LM}
\chi_{LM}^2 = \sum_k^K\lambda_k\chi_k^2,
\end{equation}
where $\lambda_k$ are positive parameters whose role is to perform the balancing, and $\chi_k^2$ is the usual model $\chi^2$ for an individual constraint.  The role of the $\lambda_k$ is described in more detail in \citet{Long2012}, as is the process for the determination of their values.

The second term of Eq.~(\ref{eqn:Fe}) imposes a constraint that $dS/dt$ should be zero. After differentiation it provides the weight derivative term in the weight evolution equations. 
The third term is the global regularization which we will discuss later in \S~\ref{sec:reg} for our M87 data. As the particle weights are taken as fractional luminosity, the last term of Eq.~(\ref{eqn:Fe}) (involving $\lambda_{sum}$) ensures that the total luminosity of the model does not vary. The manually tunable parameters ($\epsilon$, $\mu$, $\lambda_{sum}$, and potentially $\lambda_k$) are used to modify the behaviour of the model as appropriate to the real system and data being modelled.

In order to determine whether or not the particle weights have converged, we use the same mechanism as \citet{Long&Mao2010}. Reproduction or otherwise of the observed data is handled via $\chi^2_{LM}$.

\subsection{Discrete data}
Discrete observables (e.g. GC data) are directly involved in the model as constraints. 
The probability, $p_{D,j} = p_{D,j}(\mathbf{x}_{\perp j}, v_{\parallel j})$
of the model reproducing a discrete LOS measurement ($x_j$, $y_j$, $v_{\parallel j}$) is found by convolving the LOS velocity distribution (LOSVD) with a Gaussian incorporating the observational errors $\sigma_j$. The LOSVD is usually not analytic.
For modelling purpose, we divide the model `sky' into many bins in polar coordinates as illustrated in Fig.~\ref{fig:GCk}. 
Each discrete data point is located in one bin and, correspondingly, at any one time, there will be many particles ($x_i$, $y_i$, $v_{\parallel i}$) currently located in the bin. 

The probability $p_{D,j}$ is calculated from the particle data as
\begin{equation}
\label{eqn:PDJ}
p_{D,j} = \frac{1}{\sqrt{2\pi} \sigma_j} \frac{\sum_i^N\delta_{ij}w_i \exp \Bigr[-\frac{(v_{\Vert i} - v_{\Vert j})^2}{2\sigma_j^2}  \Bigr]}{\sum_i^N\delta_{ij}w_i},
\end{equation}
where the selection function $\delta_{ij}$ takes the value 1 if particle $i$ is located in the bin corresponding to observation $j$ and is 0 otherwise, and $N$ is the total number of test particles. For modelling the M87 GCs, we use $16\times16$ bins to reduce the size of the bins so that the model calculations are more representative of the measurements. 
The contribution to the weight evolution is found by constructing the log likelihood function $\mathcal{L}$
\begin{equation}
\label{eqn:LD}
\mathcal{L} =  \sum_j^M \ln{p_{D,j}},
\end{equation}
where $M$ is the number of discrete data points. Eq.~(\ref{eqn:Fe}) then becomes
\begin{equation}
\label{eqn:Fe2}
F = - \frac{1}{2}\chi_{LM} ^2 +  \lambda_D \mathcal{L} +  \frac{1}{\epsilon} \frac{d}{dt} S \mbox{} + \mu S - \frac{\lambda_{sum}}{2}(\sum_i^Nw_i -1)^2, 
\end{equation}
where $\lambda_D$ is another constant we need to adjust. By evolving the particle weights, we obtain the model which has the maximum likelihood of reproducing the discrete data sets and the $\chi^2$ modelled observables.

\subsection{Gravitational potential}
M87's GCs and stars are both tracers of the same gravitational potential, where the potential is a combination of a luminous matter potential, a dark matter potential and a central black hole. 

The luminous matter mass is dominated by M87 stars.  By comparison the mass of the GCs is negligible.  As indicated in \S~\ref{sec:phdata}, we assume an axisymmetric spatial distribution for the stars. An axisymmetric luminous matter potential is constructed by deprojecting the 2D stellar surface brightness profile.   
The stellar distribution of M87 is very round: the MGE fit shows that the short-to-long axis ratio is between 0.7-0.99 at different radii.

The flattening of dark matter and luminous matter are correlated, with dark matter usually being systematically rounder than luminous matter \citep{Wu2014}.  We assume a spherical dark matter potential in our models. 

\subsubsection{Luminous matter potential}
\label{sec:MGE}
An observed 2D surface brightness is the projection of a 3D luminosity density distribution. An edge-on inclination is assumed and so the deprojection is unique \citep{Gerhard&Binney1996}. Assuming axisymmetry, the stellar gravitational potential of a galaxy can be calculated by assuming (or knowing) the galaxy's inclination to the line of sight and deprojecting the multi-Gaussian expansion (MGE) of its surface brightness to give a 3D density. With an assumed constant mass-to-light ratio, Poisson's equation may then be solved to obtain the gravitational potential. 
The MGE technique is described in \cite{Emsellem1994} and applied in \cite{Cappellari2006}, and we do not repeat the detail here. For M87, we use the MGE in \cite{Cappellari2006} based on the {\it Hubble Space Telescope (HST)/WFPC2} and ground-based MDM photometry in the $I$ band. 

The technique for constructing the potential from multi-Gaussians in M2M models was implemented in \cite{Long2012}.
For M87, the surface brightness measurements extend to $R=40$ kpc where the MDM observations end.  For modelling purposes, we extrapolate the potential out to $R=80$ kpc using the MGE fit.

\subsubsection{Central supermassive black hole}
It is widely believed that a supermassive black hole resides at the centre of M87 (e.g \citealt{Young1978}; \citealt{Ford1994}). This component can be modelled by adding a point mass in the centre of the model. However, the radius of influence of the black hole is usually small. Taking M87 as an example, 
the mass of the black hole is $(6.6\pm0.4)\times10^9 M_{\odot}$ \citep{Gebhardt2011}. At a distance of $d=16.5$ Mpc, this corresponds to an influence radius of $\sim 3''$. 

For M87, there are only a few SAURON data points and \textit{no} GC data point within $3''$, and thus the black hole is not important given the data available. We aim to constrain the mass and dynamics at large radii and therefore ignore the black hole for the models which follow.

\subsubsection{Dark matter potential}
\label{sec:dark}
A previous study revealed that GC data can not distinguish significantly between a cored dark matter halo and an NFW halo, but may favour slightly a cored dark matter halo \citep{Murphy2011}. We do not have substantially more data in the inner GC region, so we do not expect a change to the results in this respect.
We utilise spherical logarithmic dark matter potentials in our M2M models.
The logarithmic model was inspired by the flat rotation curves of spiral galaxies (e.g., \citealt{Persic1996}) but is frequently used in the modelling of elliptical galaxies (e.g., \citealt{Murphy2011}). The gravitational potential is given by
\begin{equation}
\Phi(r) = \frac{V_s^2}{2}\ln (R_s^2 + r^2),
\end{equation}
where $V_s$ is the scale velocity and $R_s$ is the scale radius. The corresponding density profile is
\begin{equation}
\rho(r) = \frac{V_s^2(3 R_s^2 + r^2)}{4 \pi G (R_s^2 + r^2)^2},
\end{equation}
and the mass profile is
\begin{equation}
\label{eqn:logmass}
M(<r) = \frac{1}{G} \frac{V_s^2 r^3}{R_s^2 + r^2}.
\end{equation}
Unlike the density cusp at the centre of the NFW model \citep{Navarro1996}, there is a core at the centre of the logarithmic model where the density approaches a constant. 
When $r>>R_s$, the density decreases as $r^{-2}$. The logarithmic model maximizes the stellar contribution to the mass in the central regions, and thus can be used to create a `minimal' dark matter halo scenario.
 
\subsection{Parameter estimation}
\label{sec:paramest}
In our M87 M2M models using a luminous matter potential and a logarithmic dark matter halo, there are three, free parameters to be estimated, the stellar mass-to-light ratio $M/L$ and the dark matter scale velocity $V_s$ and scale radius $R_s$.  In this section we explain how these parameters, which we will refer to as $\mathbf{p}$, will be estimated.

We base our approach on \citet{Morganti&Gerhard2013} who argue that it is not appropriate to apply `normal' $\chi^2$ analyses to M2M models.  In the standard $\chi^2$ procedure, one uses certain $\Delta \chi^2$ values for given degrees of freedom to determine the confidence levels of parameters. \citet{Morganti&Gerhard2013} show that the assumptions underlying such an approach are not met in M2M models (see their \S~4).  To assist us in finding the best fitting parameter values, we define a new function $G(\mathbf{p})$
\begin{equation}
\label{eqn:Gdef}
G(\mathbf{p}) \equiv -\frac{1}{2} \chi^2_{LM} + \lambda_D \mathcal{L}.
\end{equation}
For a given model, $G$ is only calculated once, at the end of the modelling run. From equation~(\ref{eqn:Fe}), it can be seen that $G(\mathbf{p})$ is in fact the first 2 terms of $F$ and is negative.  
Our $G(\mathbf{p})$ differs from that in \citet{Morganti&Gerhard2013} in two respects.  Firstly it is the negative of \citeauthor{Morganti&Gerhard2013}'s, and secondly our $\chi^2_{LM}$ is defined as in equation~(\ref{eqn:chi2LM}) and contains the $\lambda _k$ parameters. \citet{Morganti&Gerhard2013} omit their equivalent of the $\lambda _k$ parameters.  The first change is minor and reflects our personal preference.  The second change, which we regard as essential in maintaining the mathematical consistency of our overall M2M method, introduces a more general matter with M2M methods.

A key issue that all users of M2M methods based on \citet{Syer1996} have to resolve is how to balance numerically the weight evolution equation (section \ref{sec:m2m}) such that no single observable dominates the determination of
the particle weights.  de Lorenzi et al., Long \& Mao, Hunt \& Kawata, Morganti et al. all have different approaches.    A second related issue is whether or not whatever mechanism is used to achieve the numerical balance should also 
be part of the 'merit function' used to compare models.  As is clear, our preference is for mathematical consistency of approach and this continues the approach taken in the earlier Long \& Mao papers.  Other researchers have adopted alternative strategies.  

For a set of modelling runs spanning some region in $\mathbf{p}$, we use $\Delta G$ defined as
\begin{equation}
\Delta G = G_{max} - G,
\end{equation}
where $G_{max}$ is the maximum value of the end of run $G$s within the set.  

\citet{Morganti&Gerhard2013} set parameter confidence limits for their NGC 4494 models using $\Delta G$ values determined from the probability distribution function produced from Monte Carlo simulations of models with mock data. They find, as well as other results, that the confidence limits obtained from the simulations are consistent with the fluctuations of $\Delta G$ near its minimum (see their \S~6.1). We assume that this result is not specific to their modelling and the data being modelled but is generally true, and may be applied to our M87 M2M models to perform the parameter estimation we require (see \S~\ref{sec:chiana} for the detailed calculations). As a consequence, we do not perform any Monte Carlo simulations with mock data.  Our best fitting model parameters are determined from only the models in the region of the minimum in $\Delta G$. 

For the avoidance of doubt, it is not the full method and results from \citet{Morganti&Gerhard2013} that we use but a subset.  We base our approach on theirs but, as the earlier text in this section shows, we do not follow it exactly and there are clear implementation differences, risks and assumptions.

\section{Modelling M87}
\label{sec:modelM87}
In this section, we construct and run the M2M models for M87. We use only a logarithmic dark matter halo plus a luminous matter gravitational potential constructed by deprojecting the 2D surface brightness using a MGE. A central black hole is not included in the modelling. From the potentials, we have three free parameters: the stellar mass-to-light ratio $M/L$, and the dark matter scale radius $R_s$ and scale velocity $V_s$.

Ideally, the free parameters should be modelled simultaneously, and we would use the combined $\chi^2$ for the SAURON data and $\mathcal{L}$ from modelling GCs to determine the best fitting model parameters. However, for M87, the GCs and the SAURON data will be modelled separately for the reasons noted in \S~\ref{sec:phdata}. The two sets of models will utilise the same parameter grid.

The GCs provide good constraints on the dark matter distribution, but weak constraints on the stellar $M/L$. Conversely, the SAURON data provide good constraints on the stellar $M/L$, but nearly no constraints on the dark matter distribution. Consequently we only apply a subsection of the parameter grid to the SAURON data. The modelling steps are as follows.
\begin{enumerate}
\item \label{pt:GC}GCs are modelled first with a parameter grid of $14 \times 13 \times 6$ on $V_s \times R_s \times M/L$ with\\
$V_s = (3.5,4.0,4.2,4.4,4.6,4.8,5.0,5.2,5.4,5.6,5.8,$ $6.0,6.2,6.4) \times 124.5$ km s$^{-1}$, \\
$R_s = (0.3,0.5,0.6,0.7,0.8,0.9,1.0,1.1,1.2,1.3,1.4,$ $1.5,1.7) \times 40.8$ kpc, and \\
$M/L_{\mathrm I} = (4.9, 5.39, 5.88, 6.37, 6.86, 7.35) $ $M_{\odot}/L_{\odot}$.\\
We find that $V_s$ and $R_s$ are constrained well by the kinematics of the GCs. The best fitting parameters we obtain are $V_s \sim 600$ km s$^{-1}$, and $R_s \sim 40$ kpc. 
\item \label{pt:sauron}We then model the SAURON data on a subsection of the GC parameter grid with\\ 
$V_s = (4.0,4.2,4.4,4.6,4.8,5.0,5.2,5.4) \times 124.5$ km s$^{-1}$, \\
$R_s = (0.7,0.8,0.9,1.0,1.1,1.2,1.3,1.4,1.5) \times 40.8$ kpc, and \\
$M/L_{\mathrm I} = (4.9, 5.39, 5.88, 6.37, 6.86, 7.35) $ $M_{\odot}/L_{\odot}$ .\\
\item Finally, the model parameters and their uncertainties are determined from the combination of these two sets of models using the approach in \S~\ref{sec:paramest}. Because the constraints are different in the GC and SAURON models,  $G$ and in particular $\Delta G$ also differ.  We refer to the $\Delta G$s as $\Delta G_{\mathrm{GC}}$ and $\Delta G_{\mathrm{SN}}$ respectively.
\end{enumerate}

We monitor convergence of the particle weights (see \S~\ref{sec:m2m}) over the last 20 half mass dynamical time units of the modelling runs using a convergence tolerance of $5\%$.  All our models have more than $95\%$ of the particle weights converged so we consider particle weight convergence to be a non-issue for our work.

Before applying the M2M method to M87 and its GCs, we test our procedure for modelling discrete data using NGC 4374 planetary nebulae data (see appendix \ref{app:4374}). We find our procedure behaves satisfactorily.

\subsection{Particle initial conditions}
\label{sec:ini}
We set up our initial M87 models with $N = 500000$ particles extending to $\sim 800$ kpc, which is larger than the region for which we have data constraints in order to ensure we have an appropriate number of higher velocity particles.
All particles are given the same initial weight and prior equal to $1/N$.

The distribution functions of the GCs and the stars are different and thus we create separate M2M models for the GCs and stars. For the M87 stellar models using SAURON data, the particle initial conditions are as described in \cite{Long2012}.  For the GC models, the initial spatial distribution of particles matches the GC number density profile.  Velocities are first set using the velocity dispersion function obtained from the isotropic Jeans equation using the GC number density and the potential to be modelled. The potential includes contributions from both dark matter and luminous matter.
A small rotation is added for specified fractions of particles at particular radii to assist the M2M modelling in matching the observed GC dynamics (as discussed in \S\ref{sec:GCki}). Adding rotation modifies the particles' velocities causing the velocity distribution to become anisotropic.  Thus, neither the stellar models nor the GC models have isotropic velocity dispersions before M2M modelling commences.

Since we are running M2M models with free parameters in the potential, we create a particle system for each potential as required. This ensures that the particles' velocities match the potential initially, and we only need to match the observables by adapting the weights during the modelling.

\subsection{Regularization}
\label{sec:reg}
The M2M approach to dynamical modelling is an ill-posed problem in that there are many more weights to be determined than there are observational constraints. 
\cite{Long2013} identify some of the points to be considered in deciding whether or not to use regularization with the M2M method. In our case, the M87 GC data show kinematic fluctuations at certain radii, and some of the data points are of uncertain quality. We do not want to over-fit the data and so we choose to use entropy regularization. 

The amount of regularization is controlled by the parameter $\mu$ which is used to achieve a desired balance between fitting the observed data and delivering a smooth solution. 
Entropy based regularization prevents a particle weight $w_i$ moving too far away from its `prior' weight $m_i$ ($m_i/e$ to be precise). Too much regularization may bias the model fit to the observational data. 

\begin{figure}
\centering\includegraphics[width=7.5 cm]{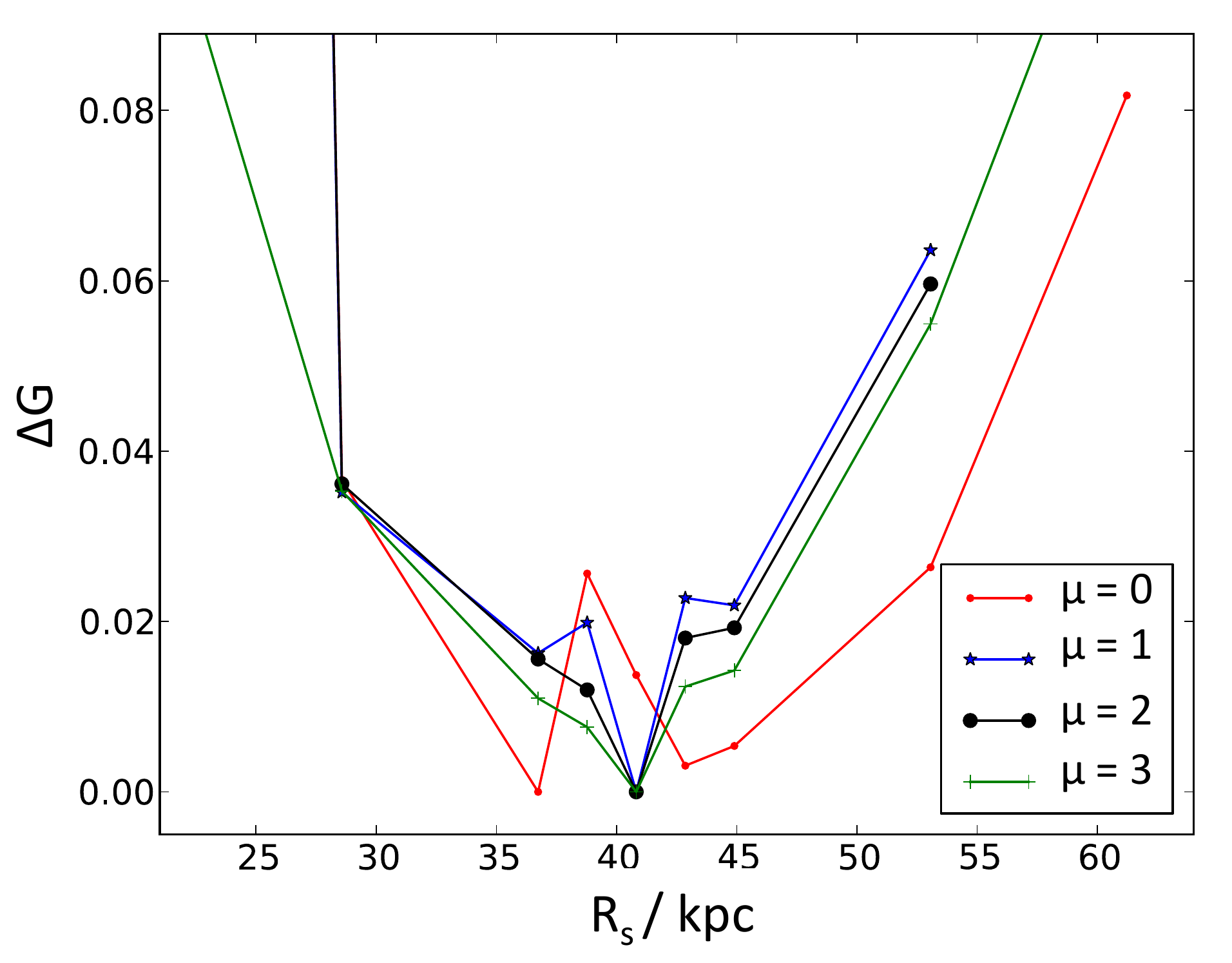}
\caption{Test of regularization. The red, blue, black and green lines indicate the four sets of models with $\mu = 0,1,2,3$ respectively.}
\label{fig:regu}
\end{figure}

To examine how regularization actually affects the determination of our model parameters, we run a series of controlled tests. We deliberately choose a parameter region where a minimum is known to exist to help us assess the impact of regularization. As anticipated, the smoothing effect of regularisation in this case does make the minimum more apparent.

Different values of $\mu$ are used in a test M2M model (utilising our M87 GC data) in which we fix two of our three, free parameters (mass-to-light ratio $M/L_{\mathrm I} = 5.88 $ $M_{\odot}/L_{\odot}$, dark matter scale velocity $V_s = 625$ km s$^{-1}$) and then vary the third, the dark matter scale radius $R_s$.
Fig.~\ref{fig:regu} shows how $\Delta G_{\mathrm{GC}}$ varies as $R_s$ is varied.
The red, blue, black and green lines represent the models with $\mu =0,1,2,3$ respectively. 

We use the minimum $\Delta G_{\mathrm{GC}}$ of each set of models to indicate the best fit.
For the $\mu = 0$ set of models (no regularization), $\Delta G_{\mathrm{GC}}$ (red line) shows fluctuations, and it is hard to determine the best fitting value of $R_s$. The other three sets of models reach their minimum $\Delta G_{\mathrm{GC}}$ value at the same value of $R_s = 40$ kpc. The $\mu = 1$ set of models (blue line) still has some fluctuations around its minimum. 
The GC kinematic data is noisy so one needs to avoid over-fitting to the noise level. However, when $\mu$ is large, it prevents fitting to the data, so the difference between models becomes smaller, e.g., the green line ($\mu=3$) is flatter than the black line ($\mu=2$) around the minimum. We need sufficient difference to distinguish different models, so $\mu = 2$ appears to be a reasonable choice for the GC model of M87 and that is what we use.

For consistency with \citet{Long2012}, we use no regularization ($\mu = 0$) when modelling the SAURON data.

\subsection{Modelling GC kinematics}
\label{sec:MGC}
The surface number density profile (with an assumed relative measurement error of $10\%$) and the discrete GC LOS velocities are used as M2M model constraints. For the GC models, $G(\mathbf{p})$ (see equation~\ref{eqn:Gdef}) may be simplified to 
\begin{equation}
G(\mathbf{p}) = -\frac{1}{2} \lambda_{sd} \chi^2_{sd} + \lambda_D \mathcal{L}.
\end{equation}
We take $\lambda_{sd} = 8\times10^{-3}$ and $\lambda_D = 4\times10^{-3}$.  The log likelihood function $\mathcal{L}$ for the GCs' discrete velocities is calculated using equation~(\ref{eqn:LD}). 

\begin{figure}
\centering\includegraphics[width=\hsize]{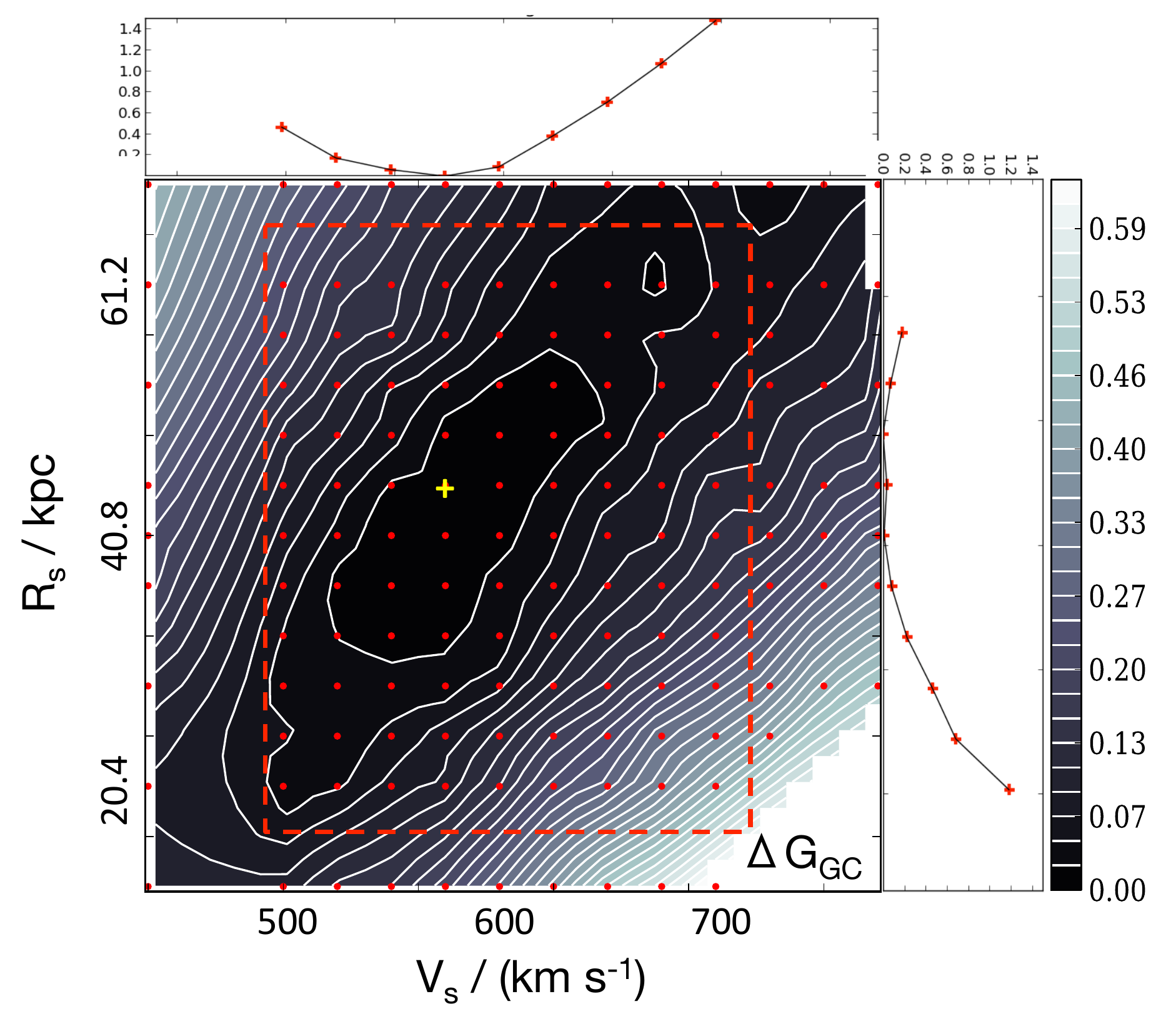}
\caption{Constraints on $V_s$ and $R_s$ for $M/L_{\mathrm I} = 6.37 $ $M_{\odot}/L_{\odot}$. The dots indicates the parameter grid points we model, and the colour bar shows the value of $\Delta G_{\mathrm {GC}}$. The data points have not been smoothed but have been interpolated for plotting purposes. The yellow plus represents the best-fitting model obtained by model marginalisation on $V_s$ and $R_s$ as shown in the side panels. 
The red rectangle indicates the grid points we model with the SAURON data.}
\medskip
\label{fig:conM87}
\end{figure}

The parameter space for $\mathbf{p}$ is as in \S~\ref{sec:modelM87} point \ref{pt:GC} and we run a M2M model for each point in the space.  We obtain a region with minimum $\Delta G_{\mathrm {GC}}$ on the parameter space, as illustrated in Fig.~\ref{fig:conM87} for $M/L_{\mathrm I} = 6.37 $ $M_{\odot}/L_{\odot}$.  The dots indicates the $V_s \times R_s$ grid we have run. In total, we have run five sets of such models with different $M/L_{\mathrm I}$.  The square indicates the subset of parameter values we use subsequently with the SAURON data. 

The constraints on $V_s$ and $R_s$ are strong, although $V_s$ and $R_s$ are degenerate diagonally. 
Marginalization gives a best fitting model with $V_s \sim 573$ km s$^{-1}$ and $R_s \sim 48.9$ kpc for the set with $M/L_{\mathrm I} = 6.37 $ $M_{\odot}/L_{\odot}$.  We look to refine these parameter values by performing further $\Delta G$ analyses (see \S~\ref{sec:chiana}).

The GC kinematics have little constraining influence on the stellar mass-to-light ratio $M/L_{\mathrm I}$. The set of models with different $M/L_{\mathrm I}$ yield a minimum $\Delta G_{\mathrm{GC}}$ region with slightly different dark matter halo parameters. The minimum $\Delta G_{\mathrm{GC}}$ area moves diagonally with the value of $\delta V_s < 20$ km s$^{-1}$ and $\delta R_s < 10$ kpc within the $M/L_{\mathrm I}$ range we run.  A high $M/L_{\mathrm I}$ ratio leads to a dark matter halo with smaller $R_s$ and $V_s$.

\subsection{Modelling SAURON data}
For modelling with SAURON data, we use surface brightness and the Gauss-Hermite coefficients $h_1$ to $h_4$ of the LOS velocity distribution as observable constraints. The coefficients $h_5$ and $h_6$ (see \citealt{Long2012}) are not used.   The function $G(\mathbf{p})$ has the form (see equation~\ref{eqn:Gdef})
\begin{equation}
G(\mathbf{p}) = -\frac{1}{2} \left ( \lambda_{SB} \chi^2_{SB} + \sum _{i=1} ^4 \lambda_{h_i} \chi^2_{h_i} \right ).
\end{equation}
There is no discrete data log likelihood term since we are not using discrete data as a constraint.  For the $\lambda$ values, we use the same values as used in the best fitting M87 M2M model from \citet{Long2012}.  That is $\lambda_{SB} = 4.0\times 10^{-3}$, $\lambda_{h_1} = 2.08 \times 10^{-5}$, $\lambda_{h_2} = 3.94 \times 10^{-5}$, $\lambda_{h_3} = 2.04\times10^{-5}$ and $\lambda_{h_4} = 4.12\times10^{-5}$.

The parameter space for $\mathbf{p}$ is as in \S~\ref{sec:modelM87} point \ref{pt:sauron} (the same subset of the parameter space described in \S~\ref{sec:MGC}) and we run a M2M model for each point in the space.  From the models we construct $\Delta G_{\mathrm{SN}}$ to assist in the parameter estimation.  In addition, we also run a set of models with no dark matter to repeat the investigation of \cite{Long2012} and obtain good agreement with their result. 

Analysis of $\Delta G_{\mathrm{SN}}$ is covered in \S~\ref{sec:chiana}.

\subsection{Parameter estimation of M87}
\label{sec:chiana}

An analysis based on the $\Delta G$ values is performed to estimate both the best-fitting modelling parameters and their uncertainties.  Fig.~\ref{fig:chi2} shows the $\Delta G$ values plotted against the three modelling parameters. The red dots represent $\Delta G_{\mathrm {GC}}$, and black stars represent $\Delta G_{\mathrm {SN}}$. We note three things from Fig.~\ref{fig:chi2}:

\begin{figure}
\centering\includegraphics[width=\hsize]{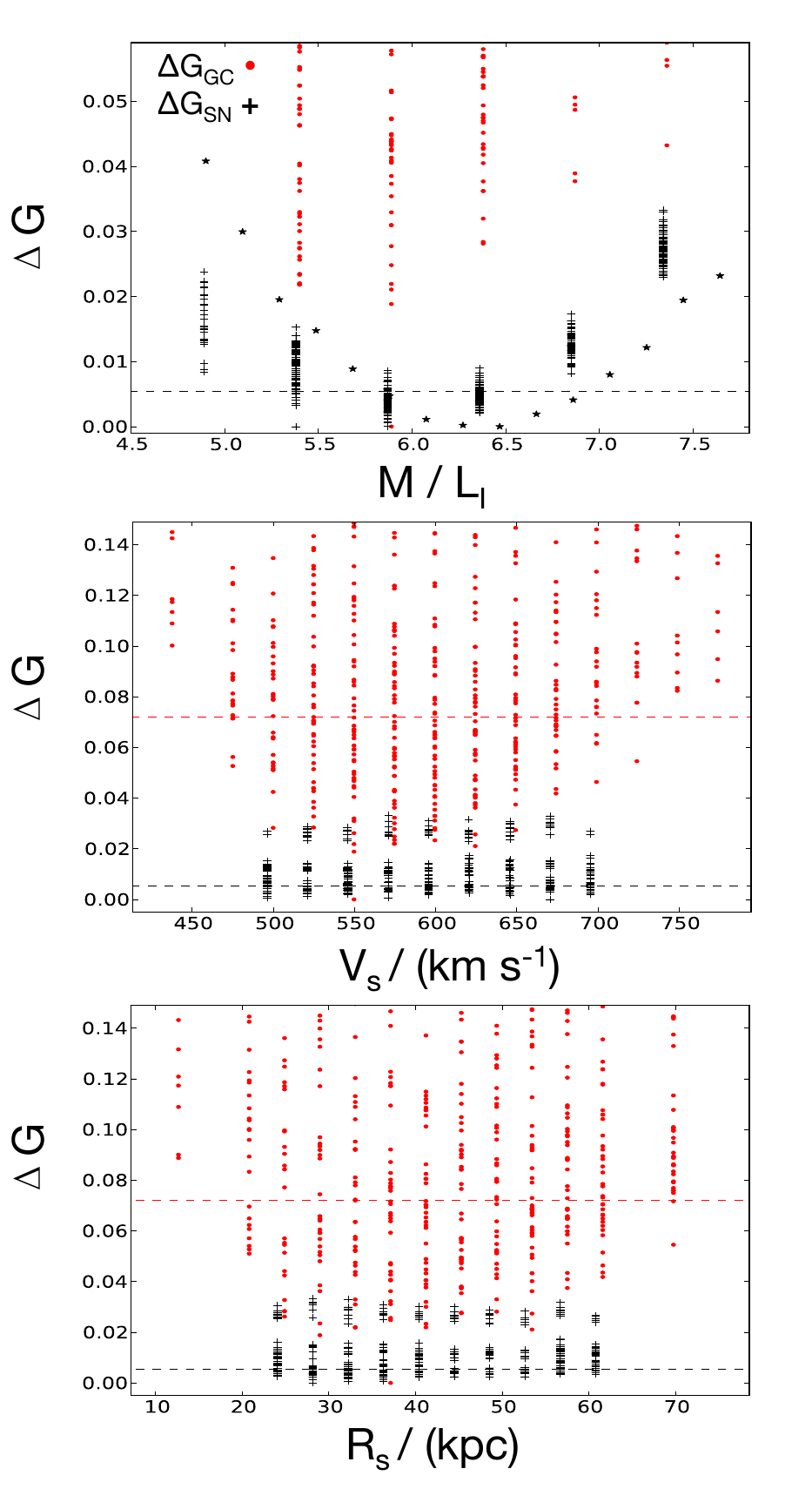}
\caption{$\Delta G$ values plotted against the three modelling parameters. The black pluses represent $\Delta G_{\mathrm{SN}}$  from the subset parameter grid model runs while the red dots represent $\Delta G_{\mathrm{GC}}$ from the complete grid model runs. 
The black asterisks in the top panel represents $\Delta G_{\mathrm{SN}}$ of the set of models with no dark matter halo. 
The red and black dashed lines represent the $70\%$ confidence levels for $\Delta G_{\mathrm {GC}}$ and $\Delta G_{\mathrm {SN}}$. }
\medskip
\label{fig:chi2}
\end{figure}

\begin{enumerate}
\item The $\Delta G_{\mathrm{SN}}$ and $\Delta G_{\mathrm {GC}}$ values differ by approximately an order of magnitude. This is because we are using different observable constraints in the two sets of models. 
\item As shown in the top panel, $\Delta G_{\mathrm{SN}}$ values are sensitive to changing the $M/L_{\mathrm I}$ values.
We choose to determine $M/L_I$ from only $\Delta G_{\mathrm{SN}}$. 
\item There is a minimum region of $\Delta G_{\mathrm{GC}}$ when projected along the $R_s$ and $V_s$ axes.  The variation in $\Delta G_{\mathrm{SN}}$ appears to be random (i.e, the SAURON data provides nearly no constraint on the dark matter halo). We choose therefore to use only $\Delta G_{\mathrm{GC}}$ to determine $V_s$ and $R_s$.
\end{enumerate}

\begin{figure}
\centering\includegraphics[width=7.cm]{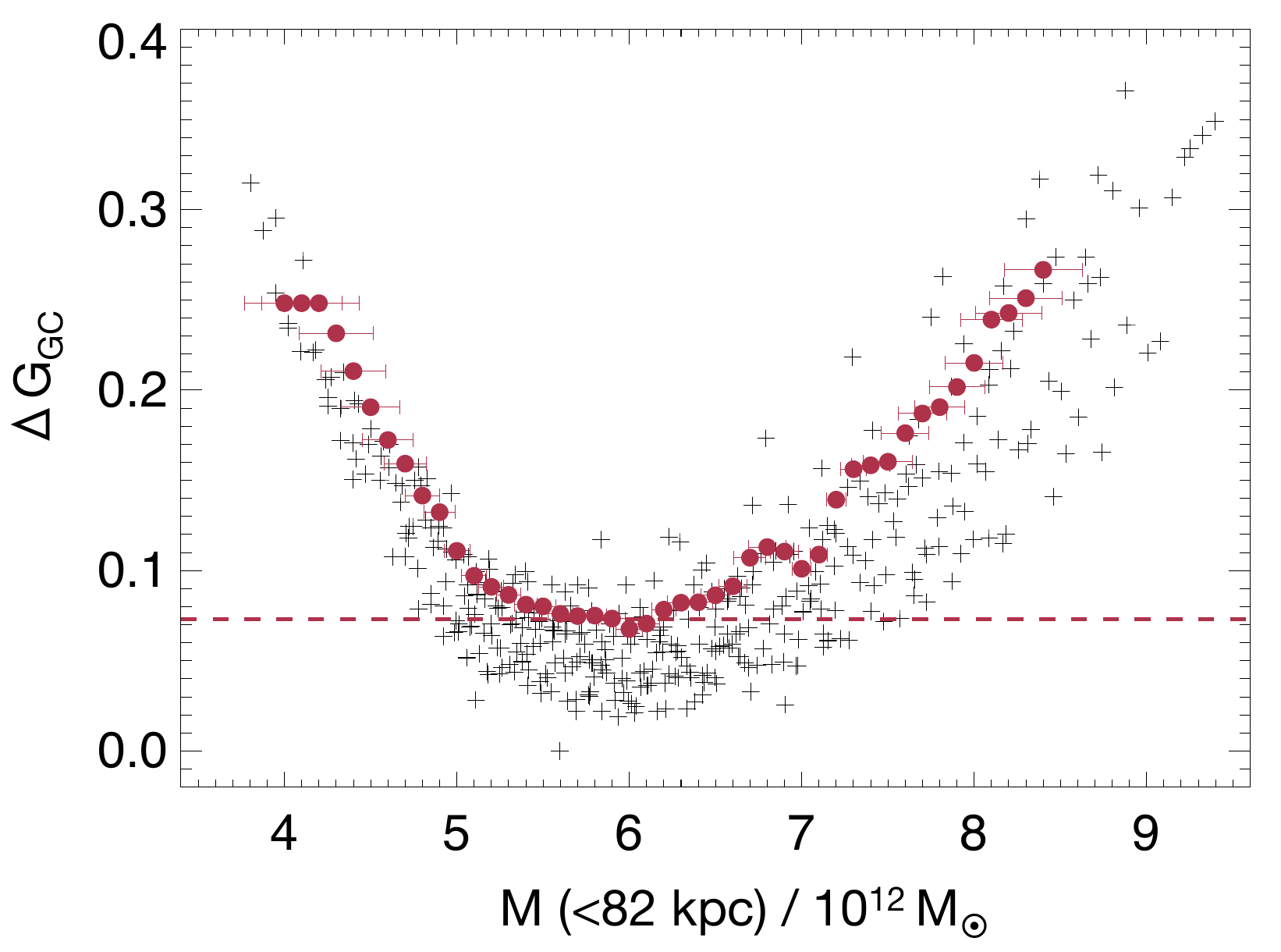}
\caption{The enclosed mass at $r=82$ kpc vs. $\Delta G_{\mathrm{GC}}$. Each point represents a M2M model. The $70\%$ confidence level from the $\Delta G_{\mathrm{GC}}$ fluctuations near the minimum is indicated by the dashed line $\Delta G_{\mathrm{GC}} = 0.073$. }
\label{fig:Ggc}
\end{figure}

As indicated in section \ref{sec:paramest}, we estimate the confidence regions for the parameters from the $\Delta G$ fluctuations near the minimum.
Since the total mass within a given radius is influenced by all three parameters, we use it to determine our confidence limit $\Delta G$ values.  In the inner part, the mass is dominated by the stellar mass-to-light ratio, and in the outer part by the dark matter halo parameters $V_s$ and $R_s$. 
For the halo parameters, we first calculate the enclosed mass at $r = 82$ kpc for each GC model (see Fig.~\ref{fig:Ggc}). The minimum  $\Delta G_{\mathrm {GC}}$ value appears to be at $M \sim 6.0\times10^{12} M_{\odot}$.  It is less clear however just how wide (in mass) the minimum region is for the purposes of selecting $\Delta G$ fluctuations near the minimum.  To clarify the width, we select a set of mass points, with a regular interval, from the overall mass range, and determine the mean and dispersion ($\sigma$) of $\Delta G$ from the nearest 30 (mass, $\Delta G$) points. For each mass point, we plot the mean value plus $1\;\sigma$ in Fig.~\ref{fig:Ggc} (the red diamonds). From the curve formed by the red diamonds, it is now considerably clearer where the minimum region is and what value of $\Delta G$ should be taken as a confidence limit (represented by the horizontal dashed line in Fig.~\ref{fig:Ggc}).  
We take as that limit $\Delta G_{\mathrm{GC}} = 0.073$ to represent a confidence level of $70\%$. We have applied our approach to the data in \citet{Morganti&Gerhard2013} Fig.~11 and arrive at the same value of $\Delta G$ as they obtained via their cumulative distribution function, that is $\Delta G = 26$.

Similarly, we calculate the enclosed total mass at $r=2.04$ kpc for each SAURON model,  
and find the minimum region in $\Delta G_{\mathrm {SN}}$ at $M \sim 9.2\times10^{10} M_{\odot}$, and then calculate the mean and dispersion ($\sigma$) of $\Delta G_{\mathrm {SN}}$ for this group of models. We take the mean value plus $1\;\sigma$ as the $\Delta G_{\mathrm {SN}}$ fluctuation near the minimum, giving $\Delta G_{\mathrm SN} = 0.0054$ as the $70\%$ confidence level. 

Based on the discussion above, we estimate the model parameters and their uncertainties (summarised in Table~\ref{tab:mparam}) as follows: 
\begin{enumerate}
\item For $M/L_I$, we choose the SAURON models with $\Delta G_{\mathrm{SN}} < 0.0054$. Note that the $V_s$ and $R_s$ values of these models are within the $1\sigma$ confidence of the GC model. For the models, we calculate the mean value and dispersion of the $M/L_I$ parameter. We take these values as the parameter estimate and its error.
We obtain $M/L_{\mathrm I} = 6.0 \pm 0.3 $ $M_{\odot}/L_{\odot}$. The set of models with no dark matter halo yields $M/L_I = 6.4\pm0.3 $ $M_{\odot}/L_{\odot}$, in agreement with \cite{Long2012}. 
\item  For $V_s$ and $R_s$, we first select the GC models with $\Delta G_{\mathrm{GC}} < 0.073$, and choose only those with $M/L_I = (5.39, 5.88, 6.37) $ $M_{\odot}/L_{\odot}$. This is because the other values of $M/L_I = (4.9, 6.86, 7.35) $ $M_{\odot}/L_{\odot}$ are excluded by the $1\sigma$ confidence level on $M/L_{\mathrm I}$.
 For these models, we calculate the mean and dispersion of the dark matter parameters, giving $V_s = 591\pm50$ km s$^{-1}$ and $R_s = 42 \pm 10$ kpc. 
\end{enumerate}
It must be remembered that we are not using parameter probability distributions with their associated confidence levels (constructed from Monte Carlo simulations, and marginalised). As a consequence, we do not wish to use unwarranted precision in stating error values and choose to quote them to one significant figure only.
 
\begin{table}
\caption{The final model parameters of M87.}
\label{tab:mparam}
\begin{tabular}{cccc}
\hline
\hline
$M/L_I$ [$M_{\odot}/L_{\odot}$]& $V_s$ [km s$^{-1}$]  & $R_s$ [kpc]& $M(<180$ kpc) [$M_{\odot}]$\\
$6.0\pm0.3$ & $591\pm50$ & $42\pm10$& $(1.5\pm0.2)\times10^{13}$\\
\hline
\hline
\end{tabular}
\end{table}

\subsection{GC data reproduction}
\label{sec:GC_bin}
To evaluate how well the GC kinematic data has been reproduced by our models, we pick the M2M model closest to the parameter values determined in \S~\ref{sec:chiana} and perform the following evaluation.

We divide the symmetrised observed data and the model's end of run weighted particle data into $6\times8$ spatial bins on the projection $R \times \phi$ plane, and then compare the velocity distribution of the data and the model in each bin. The velocity distribution of the data in each bin is not a pure Gaussian profile. Our model however does match the non-Gaussian features well. For quantitative comparison purposes, we only extract the mean velocity and velocity dispersion.

\begin{figure}
\centering\includegraphics[width=8.cm, height=9 cm]{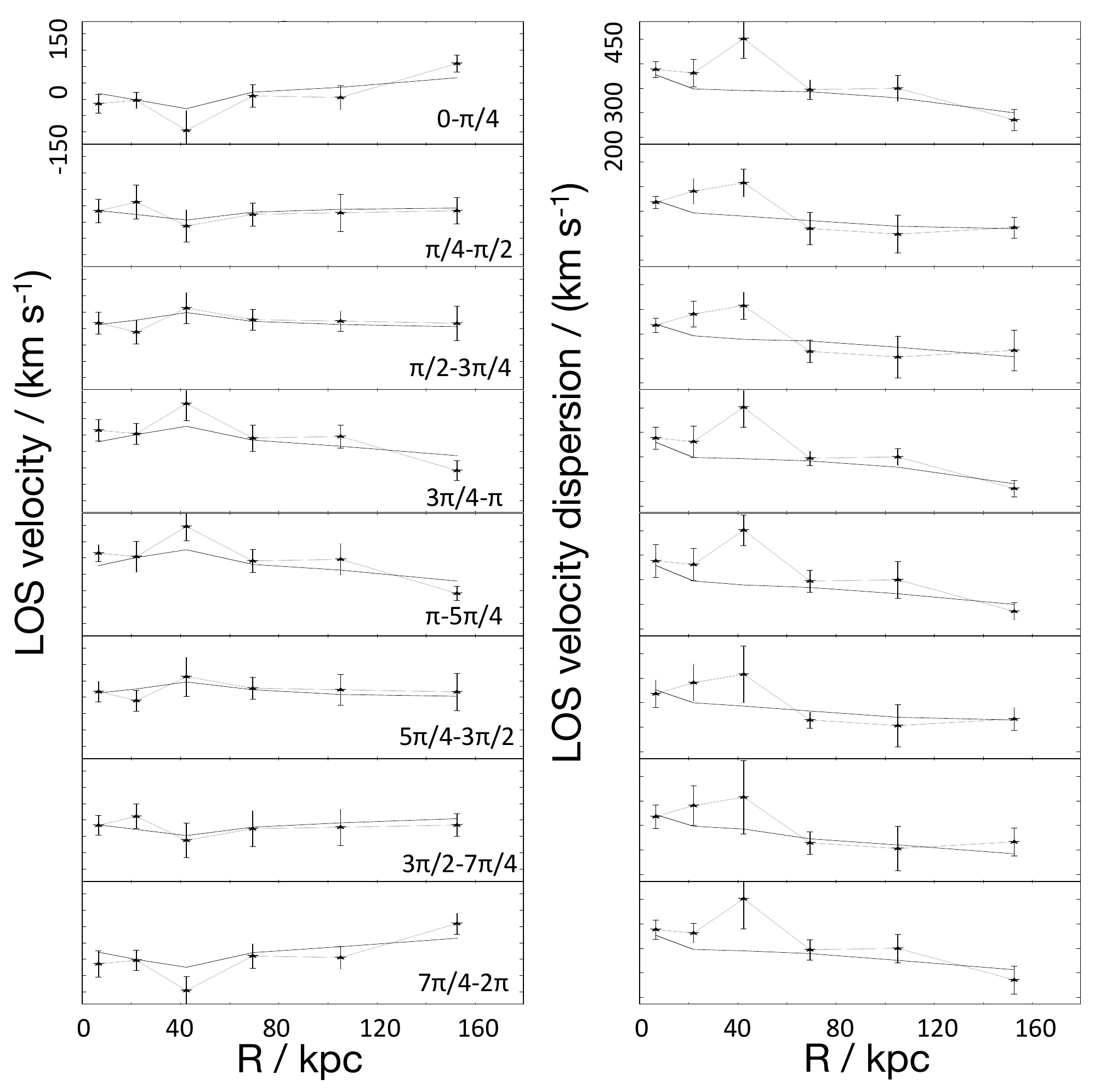}
\caption{The LOS mean velocity and velocity dispersion profiles. Each panel shows a profile along $R$ in the specified direction. The zero angle represents the direction of the major axis. The blue dots with error bar are calculated from the GC data, and the solid lines are from the model. The velocity scales are shown at the top panels.}
\label{fig:GC_bin}
\end{figure}

The resulting LOS mean velocity and velocity dispersion in each bin are shown in Fig.~\ref{fig:GC_bin}.
The left hand panels are the mean LOS velocity profile along $R$, with each panel corresponding to one $\phi$. The right hand panels are the equivalent for LOS velocity dispersion. The data values and errors shown are calculated by a bootstrap process. For each individual GC measured velocity with value $v_j$ and error $\sigma_j$, we construct a Gaussian velocity distribution with mean $v_j$ and dispersion $\sigma_j$. A new data set is produced where the velocity for each point is chosen randomly from its Gaussian distribution, and a mean velocity and velocity dispersion can then be obtained for each bin position. This process is repeated 1000 times.  We calculate the average and dispersion of the 1000 mean bin velocities as the true value of the mean velocity and its error, and similarly for velocity dispersion.

The model mean LOS velocities match the binned GC data well with a mean $\chi^2$ value of 0.5. 
For the velocity dispersion profiles (with a mean $\chi^2$ value of 0.9), the model matches the basic trend within the GC data with the model dispersion decreasing with increasing radius. The azimuthal differences in velocity dispersion are not significant: the velocity dispersion profiles at different directions are consistent within the error bar.

For the observed data at $R \sim 40$ kpc, there is a relatively large rotation along the minor axis (with a rotation velocity of 140 km s$^{-1}$), which is less well fitted by the model, and, as anticipated, the model does not fit the large velocity dispersion around this region. This dispersion can not readily be reproduced by the potential, and we deliberately employed regularization to prevent the model from attempting to match this fluctuation. We also reduced the impact of the data around 40 kpc on the model (see \S~\ref{sec:GCki}) by increasing the errors in this part of data. Having done this, the $1\sigma$ lower limit of the velocity dispersion at $R\sim 40$ kpc can marginally match our model if the dispersion gradually decreases from the centre to the outer part.  The dark matter parameters are constrained primarily by GCs at radii $>40$ kpc, thus giving confidence in the parameters derived from our modelling. 
 
\begin{figure}
\centering\includegraphics[width=8.cm, height=6.6cm]{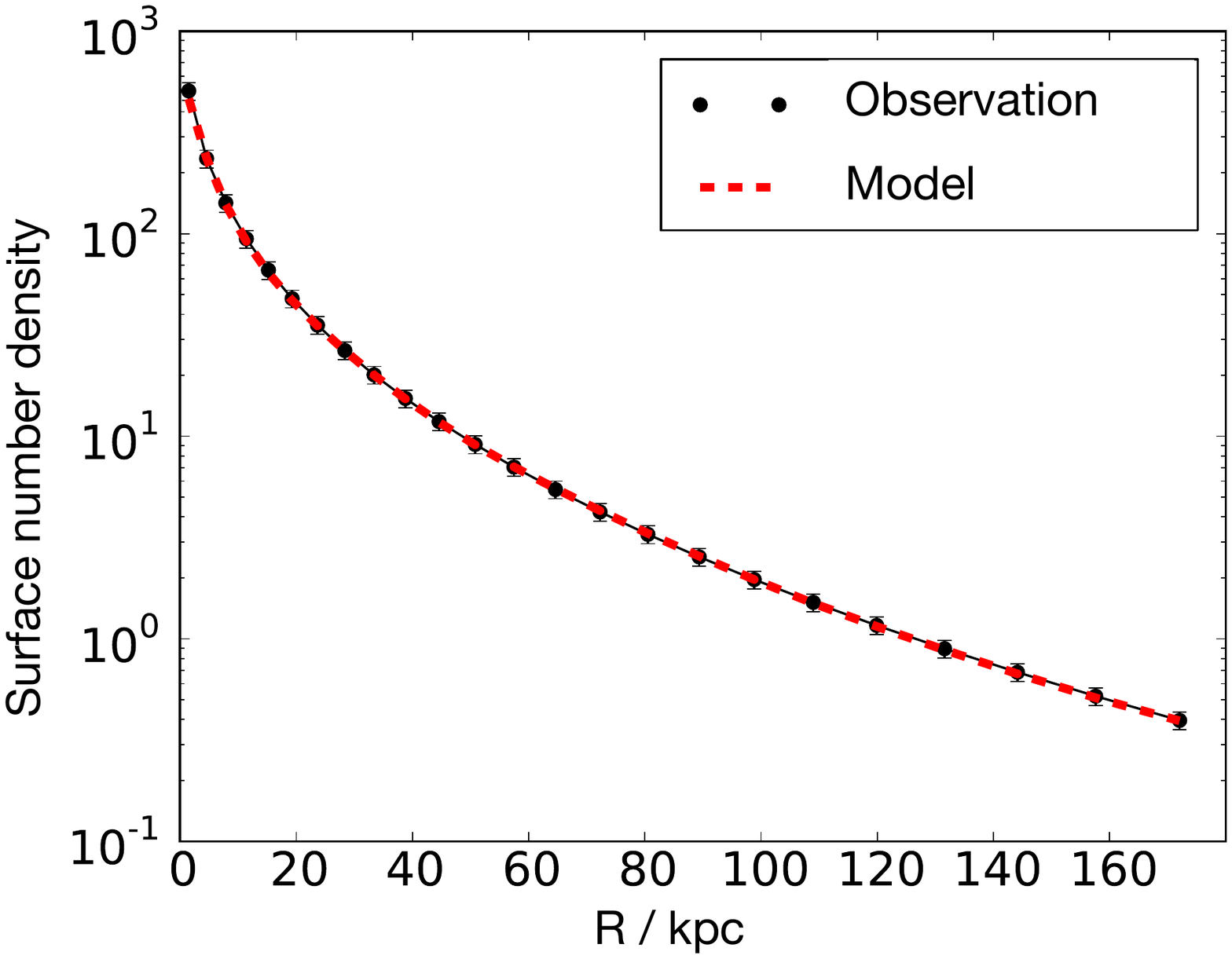}
\caption{Reproduction of the GC surface number density per kpc$^2$ for $R<180$ kpc. 24 radial bins have been used.  Normalization of the observed surface number density profile has been adapted to match the number of particles in the model.}
\label{fig:sb_gc}
\end{figure}

For completeness, Fig.~\ref{fig:sb_gc} shows how well the surface number density has been reproduced (mean $\chi^2$ value of 0.05). 

\section{Discussion}\label{sec:discuss}
\subsection{Mass distribution}
We calculate the total mass as the sum of the luminous matter and dark matter contributions.  The dark matter mass within radius is determined with equation~(\ref{eqn:logmass}). For luminous matter, the mass is determined as the stellar mass-to-light ratio multiplied by the integral of the MGE luminosity density (see \citealt{Emsellem1994}).

We find that the total mass of M87 within 180 kpc is ($1.5\pm0.2)\times10^{13} M_{\odot}$. Furthermore, the dark matter distribution is consistent with a smooth halo within 180 kpc. 
 
\begin{figure*}
\centering\includegraphics[width=\hsize]{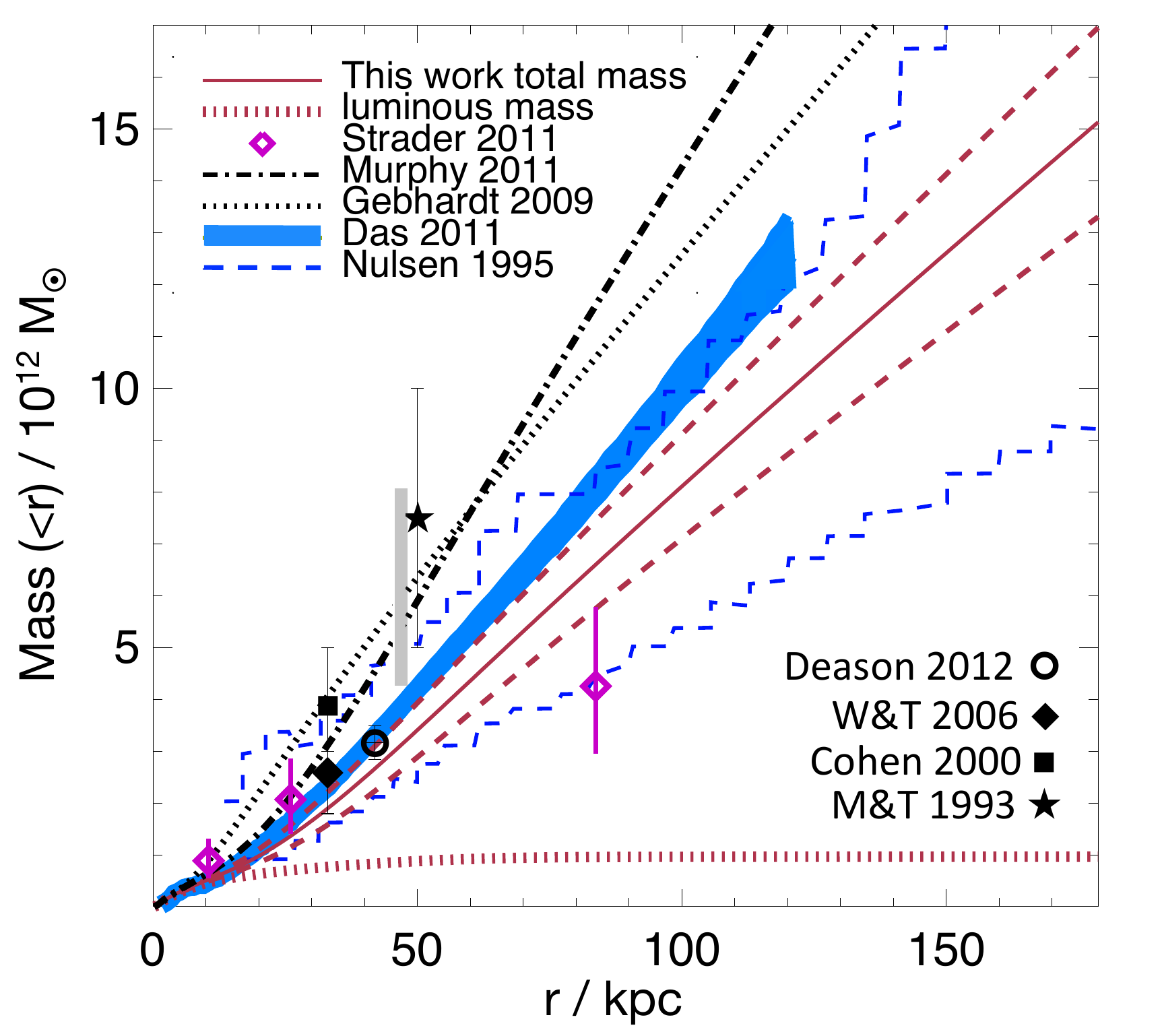}
\caption{The mass profile of M87. 
The red solid and dashed lines indicate our total mass and its uncertainties.  Our luminous mass profile is plotted with red dotted line, while the width represents its uncertainty.
The black dash-dotted line is the mass profile from M2011 and was constrained from VIRUS-p data extending to $\sim 19$ kpc and GCs extending to 45 kpc. The black dotted line is the mass profile from G2009 and was constructed using SAURON data and the same GCs as M2011. The vertical grey shade over the G2009 and M2011 profile indicates the position $r=45$ kpc, and both the M2011 and G2009 mass profiles outside of this position are extrapolated from their models. The magenta `$\diamond$'s indicate the mass obtained by \protect\cite{Strader2011} by Jeans analysis using their new GCs combining the old GCs in literatures.
The blue solid region is the mass profile and corresponding uncertainties derived from X-ray gas kinematics based on \emph{Chandra} and \emph{XMM-Newton} observations\protect\citep{Das2010}, while the region within the dark blue dashed lines is the mass and corresponding uncertainties from the early \emph{ROSAT} observation \citep{Nulsen1995}.
The black circle, diamond, square and star with error bars are the masses derived with different GC samples or different methods by \protect\cite{Deason2012}, \protect\cite{Wu&Tremaine2006}, \protect\cite{Cohen2000} and \protect\cite{MerrittTremblay1993}, respectively.  
}
\label{fig:mass-M87}
\end{figure*}

As one of the most studied galaxies, the mass of M87 has been derived many times and the literature shows a broad variation in its value. Below we compare our results with earlier studies using mass estimates based mainly on GC kinematics and X-ray studies extending to large radii. We display our comparison in Fig.~\ref{fig:mass-M87}. 

Previous mass estimates obtained using GCs kinematics are marginally consistent with our result but, with one exception, systematically larger. It must be pointed out that none of the existing studies included measurements beyond 50 kpc. Extending to 180 kpc, our data is therefore far better suited to constrain the dark matter distribution. 

Of the existing measurements, the mass measured by \cite{Deason2012} within 42 kpc is consistent with our result. However, their study assumed a constant velocity dispersion
anisotropy (see equation \ref{eqn:beta})  and they found $\beta=-1.1$,
which is inconsistent with our result (see \S\ref{sec:anisotropy}). The masses derived in earlier
studies by \cite{Wu&Tremaine2006} and \cite{Cohen2000},
based on the same data as \cite{Deason2012}, are also consistent with our mass. An even earlier study by \cite{MerrittTremblay1993} used only 43 GC velocity measurements and their
result (with its large error bar) agrees with ours to within
$1.5\sigma$. These previous results employ various assumptions and confusingly relate the GC and star surface number densities which may affect the mass estimation. We focus our comparison therefore on two recent Schwarzschild models (G2009 and M2011).

The M2011 study used not only GCs extending to 45 kpc but also VIRUS-p data out to $\sim19$ kpc, while the G2009 work was constructed using SAURON data and the same GCs as in M2011, but also used high spatial resolution long-slit observations of stellar light \citep{vanderMarel1994}.
Both M2011 and G2009 obtained larger masses than our study. For $r < 40$ kpc, their mass is $\sim 60\%$ larger than our value. 
Their GC data only extended to 45 kpc and around this radius the GCs show a large velocity dispersion which may be due to GC kinematical substructures (see Fig.~\ref{fig:GCk}). The large dispersion around 40 kpc is not consistent with the gradually decreasing velocity dispersion at larger radii ($R>60$ kpc). We choose not to fit the large dispersion in the M2M models as indicated in Fig.~\ref{fig:GC_bin}. Due to the lack of data outside of 45 kpc, the M2011 and G2009 masses are likely over-estimated. Also, other previous studies may have been affected by the large dispersion around 40 kpc as well.

In general, the mass derived in M2011 is closer to our results. The M2011 value is a balance between the VIRUS-p data and the GCs' larger velocity dispersion. In fact, M2011 showed that the mass derived with the VIRUS-p data alone is even smaller. Thus it appears that the VIRUS-p data extending to $238''$ ($\sim 19$ kpc) do have some ability to constrain the dark matter component of the total mass distribution.

\cite{Strader2011} derived the M87 mass by Jeans analysis using a GC catalogue they had constructed including 737 GCs extending to radii of $\sim 150$ kpc. They break the data set down into three radial bins and obtained the enclosed mass for each position. Compared with our results,
their mass in the two inner bins is about $20\%$ larger than ours, but is $25\%$ smaller for $r\sim 85$ kpc. Their results are in agreement with ours to within $\sim 1.5\sigma$.

M87 also has diffuse X-ray emission. The galaxy mass has been derived from \emph{ROSAT}, \emph{Chandra} and \emph{XMM-Newton} observations under the assumption that the hot gas is in hydrostatic equilibrium \citep{Das2010, Churazov2008, Nulsen1995}. The early result based on \emph{ROSAT} observations has a relative large uncertainty, while the \cite{Das2010} mass profile is based on data from Chandra and XMM, and uses a Bayesian approach, which is more accurate, to find the most likely mass profile.
The mass derived from X-ray observations is smaller than previous dynamical mass estimates.  
The mass estimated from \emph{ROSAT} observation is in good agreement with our mass out to 180 kpc.  The \emph{Chandra} and \emph{XMM-Newton} mass is slightly larger than ours, but their estimates are generally consistent with our estimates, to within $10\%$ ($1\:\sigma$) at $r< 30$ kpc, and about $20\%$ ($2\:\sigma$) larger within 120 kpc. There is evidence that the hot gas in M87 is disturbed, suggesting that the hydrostatic equilibrium assumption is not valid and this may affect the mass determination from X-ray emissions \citep{Churazov2008}. On the other hand, perhaps our dynamical studies may still be affected by the kinematic substructures as well. 
Overall, we consider the agreement between these two different probes (X-ray and our GC kinematics) to be satisfactory.

\subsection{Stellar mass to light ratio}
We estimate the M87 stellar mass-to-light ratio to be $M/L_{\mathrm I} = 6.0 \pm0.3 $ $M_{\odot}/L_{\odot}$. 
The stellar mass to light ratio of M87 has been assessed many times in the $V$ band. For comparison purposes, we transfer the previous estimates to I band with distance ($d=16.5$ Mpc)  using 
\begin{equation}
M/L_1 = (M/L_2) \times \frac{L_2}{L_1} \times \frac{d_2}{d_1}, 
\end{equation}
where $d_2$ and $d_1$ refer to the distances used for the evaluation of $M/L_1$ and $M/L_2$ respectively.

\cite{Long2012} estimated $M/L_{\mathrm I} = 7.05 $ $M_{\odot}/L_{\odot}$ at a distance $d =15.6$ Mpc and with no dark matter included. This corresponds to $M/L_{\mathrm I} = 6.7 $ $M_{\odot}/L_{\odot}$ at $d= 16.5$ Mpc, and we confirmed this result in \S~\ref{sec:chiana}.

Taking the V-band magnitude as $M_v = 8.23$,
M2011 obtained $M/L_{\mathrm V} = 9.1\pm0.2 $ $M_{\odot}/L_{\odot}$ at a distance $d = 17.2$ Mpc which corresponds to $M/L_{\mathrm I} = 7.5\pm0.2 $ $M_{\odot}/L_{\odot}$. G2009 gave $M/L_{\mathrm V} = 6.3\pm0.8 $ $M_{\odot}/L_{\odot}$ with $d = 17.2$ Mpc which implies $M/L_{\mathrm I} = 5.2\pm0.7 $ $M_{\odot}/L_{\odot}$. 

The M2011 $M/L_{\mathrm I}$ is larger because the velocity dispersion from the VIRUS-p data is systematically larger than that of the SAURON data \citep{Murphy2011}. The small $M/L_{\mathrm I}$ in G2009 may be caused by a model degeneracy.  They obtained a more concentrated dark matter distribution ($V_s = 715\pm15$ km s$^{-1}$, $R_s = 14\pm2$ kpc) which could act to reduce the stellar mass-to-light ratio.  

\subsection{The velocity dispersion anisotropy}
\label{sec:anisotropy}
To characterise the velocity dispersion anisotropy of our M2M models, we take the $\beta (r)$ parameter defined, in spherical coordinates $(r, \theta, \phi)$, as
\begin{equation}
\beta(r) = 1 - \frac{\sigma^2_{\theta}(r) + \sigma^2_{\phi}(r)}{2\sigma^2_r(r)}
\label{eqn:beta}
\end{equation}
where the $\sigma$ functions are the velocity dispersions in the indicated directions.  More detail on $\beta(r)$ may be found in \citet{BT2008}.

\begin{figure}
\centering\includegraphics[width=\hsize]{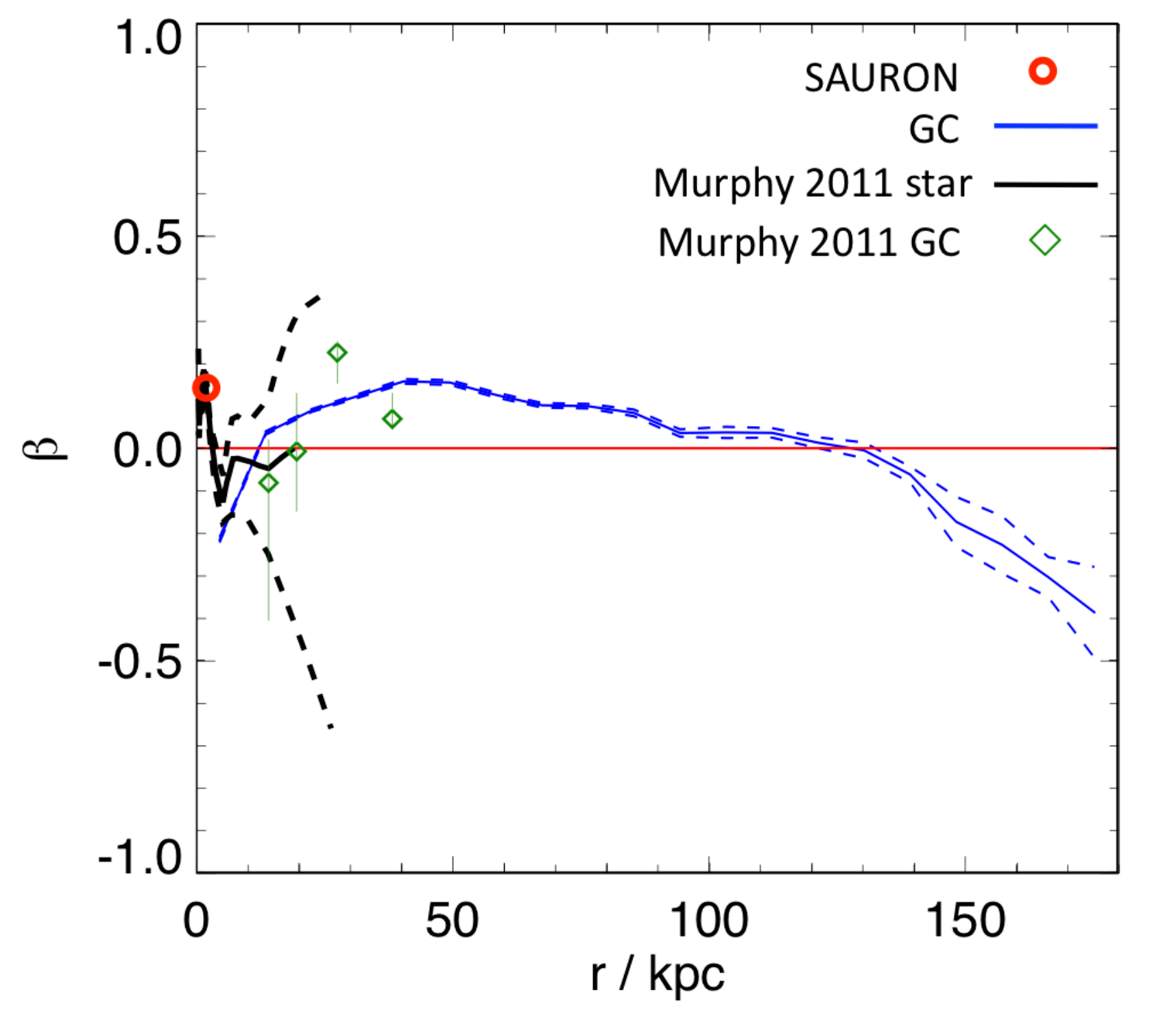}
\caption{The velocity dispersion anisotropy $\beta$ parameter as a function of radius $r$. The red circle indicates the $\beta$ value for the M2M model constrained by the SAURON data, blue solid and dashed lines are the $\beta$ profile of our GC model and its corresponding uncertainty, black solid and dashed lines are the stellar $\beta$ profile from M2011 and with its uncertainty, and green diamonds are from the GC models of M2011.}
\label{fig:GCbeta}
\end{figure}

M2M models can reproduce the velocity dispersion anisotropy in the discrete data. The blue solid and dashed lines in Fig.~\ref{fig:GCbeta} indicate the $\beta$ profile in the GC M2M model and its uncertainty. The uncertainty is derived from the M2M models within the confidence level the same as the models used to derive mass profile uncertainty. The uncertainty is larger at high radii due to the limited number of particles available to calculate $\beta$. 

The velocity dispersion anisotropy in the M87 GCs system was first studied by \cite{Cote2001}, and they found the GC orbit structure to be largely isotropic. Consistently, the $\beta$ parameter of our GC model is generally small, agreeing with the fact the M87 is located at the centre of Virgo cluster and may have evolved from multiple mergers resulting in generally isotropic orbital motions.
 $\beta$ is negative in the inner $\sim15$ kpc, and increases to become positive reaching a maximum at $R\sim 40$ kpc. It then drops to become negative again for $R>120$ kpc. By examining each velocity dispersion component, we find that $\sigma^2_r$ is smaller in the inner $\sim 20$ kpc compared to the velocity dispersion profile obtained from an isotropic Jeans analysis, and is mainly responsible for the negative $\beta$ in the centre.  For the region $R>20$ kpc, $\sigma^2_r$ is similar to isotropic values, with the deviation of $\beta$ from zero (isotropic) being mainly caused by the decrease of  $\sigma_{\phi}^2$ and $\sigma_{\theta}^2$ from isotropic values. These two dispersions are highest in the centre and decrease at $R>20$ kpc due to the partial regular rotation, with $\beta$ largest at $R\sim 40$ kpc where the rotation is at its greatest (see Fig~\ref{fig:GCk}).  The variation of velocity anisotropy in the inner region may be related to GC destruction processes which are supposed to have a dynamical selection effect on the survival of GCs (\citealt{Murali1997}; \citealt{McLaughlin1995}).

We overplot in Fig. \ref{fig:GCbeta} the $\beta$ profile converted from Fig. 12 of M2011. The anisotropy of the GC model in M2011 is generally consistent with our model. They obtain a maximum value of $\beta$ near 30 kpc where the rotation is largest in their dataset. The anisotropy of both M2011's models and our GC models are consistent with that of the stellar model out to $\sim 20$ kpc. The anisotropy of our GC model at $R<5$ kpc is uncertain due to few GCs being observed in the brightest part of the galaxy. M2011 find a stellar $\beta$ increasing to a positive value again near the centre of the galaxy. This is consistent with $\beta \sim 0.15$ in our M2M model constrained with the SAURON data.     

Fig.~\ref{fig:GCbeta} also shows that $\beta$ becomes negative again at $R>120$ kpc. There is however a large uncertainty in the $\beta$ value at large radii. This uncertainty may be a numerical artefact due to the finite number of particles we are using and the finite model size. 

The mean effective (half mass) radius $r_h$ of GCs is related to the two-dimensional (projected) galactocentric distance $R_{gc}$ observationally with $r_h \propto R_{gc}^{\alpha}$ (e.g. \citealt{Jorden2005}; \citealt{Spitler2006}; \citealt{Gomez2007}).  In an earlier simulation \citep{Webb2013}, the distributions of M87 GC orbits are used to explain the observed index $\alpha = 0.14$. In this case, they found that, assuming the GCs are tidally filled under an isothermal potential, a very high velocity anisotropy ($\beta \sim 0.99$) is needed to explain the $\alpha$ value for M87's GCs. This is in conflict with what we have found dynamically.
The velocity anisotropy $\beta$ profile from our dynamical modelling should put strong constraints on such simulations. Based on our M2M modelling, a high velocity anisotropy such as $\beta = 0.99$ would not be a preferred value for the M87 GC system.

\subsection{Model degeneracy}
\label{sec:deg}
This is the first time that kinematic data out to the approximate edge of a galaxy's gravitational boundary has been used to constrain its dark matter distribution. There are only about 20 elliptical galaxies with discrete data extending to $3-10 R_e$ in the literature \citep{Coccato2009}, including our test galaxy NGC 4374 as in appendix \ref{app:4374}. Based on the M2M modelling of M87, we discuss what kind of data can help remove degeneracies in dynamical models of galaxies.

Dynamical models mainly suffer from three levels of degeneracy, (1) mass-anisotropy degeneracy, 
(2) dark matter-luminous matter degeneracy and (3) degeneracy of parameters in the dark matter distribution even when the total mass is known. 

IFU data is helpful to remove the mass-anisotropy degeneracy. However such data usually only extends to $R<2R_e$. Having a large sample of discrete GC velocities at large radii, as in the current paper, helps significantly as it provides information on the line-of-sight velocity distribution and hence the orbit distribution at these radii. 
Even in a galaxy with a relatively large velocity dispersion anisotropy, discrete data works well to remove the mass-anisotropy degeneracy in a M2M model \citep{Morganti&Gerhard2013} using local regularization\citep{Morganti2012}. The global regularization we are using may prevent us from creating a strongly anisotropic model from isotropic initial conditions. Our models are however showing M87 to be a low anisotropy system with $\beta(r)$ values not inconsistent with \citet{Murphy2011}.  The total mass of the galaxy is well constrained by our GC kinematic observations, and so we do not believe the mass-anisotropy degeneracy for M87 to be significant.  

The luminous and dark matter degeneracy is hard to remove except when we can derive the stellar mass-to-light ratio from stellar population studies. What is usually assumed is a constant stellar mass-to-light ratio, and an analytic dark matter distribution. 
In this case, we can either try to constrain the stellar mass-to-light ratio from the central IFU data where dark matter contributes little, or constrain the dark matter distribution with discrete data extending to large radii.   
For the galaxies with discrete data extending to a few effective radii, the dark matter distribution is still relatively unconstrained, and therefore similarly the stellar mass-to-light ratio.
For M87, the dark matter distribution has been constrained well out to large radii, in the region where the stellar mass-to-light ratio has little influence.

\begin{figure}
\centering\includegraphics[width=\hsize]{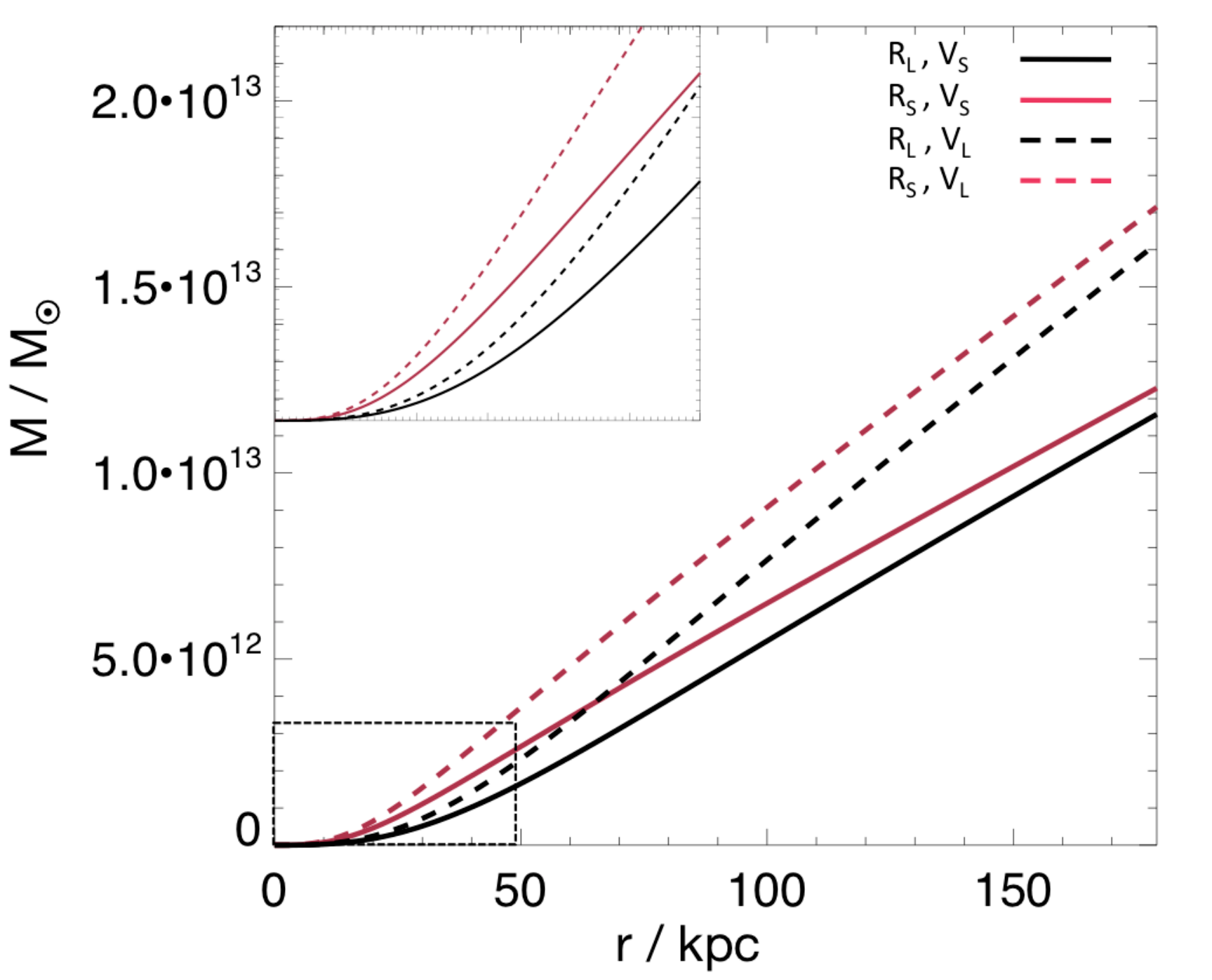}
\caption{Degeneracy of the scale velocity and scale radius in the logarithmic dark matter model. The insert panel enlarges the inner region marked by the dashed rectangle box. Each line represents a mass distribution of a logarithmic model.  Dashed lines are models with $V_{L} = 650$ km s$^{-1}$, and solid lines are models with $V_{S} = 550$ km s$^{-1}$. Black indicates $R_{L} = 53$ kpc and red indicates $R_{S} = 28.6$ kpc. (The subscript S indicates `smaller' and L is `larger'.)}
\label{fig:LOGdg}
\end{figure}

Parameters of the dark matter distribution are degenerate even when the dark matter mass within a few effective radii is constrained \citep{DeLorenzi2008, Napolitano2011}. 
When modelling M87 with GCs, we have GC kinematic data extending to $\sim 25 R_e$. As we can see in Fig.~\ref{fig:LOGdg}, the scale radius and velocity are degenerate for a logarithmic dark matter halo. The typical range for M87 solutions (with a 1.0 $\sigma$ confidence boundary) is shown with $R_{S} = 28.6$ kpc, $R_{L} = 53$ kpc, $V_{S} = 550$ km s$^{-1}$, $V_{L} = 650$ km s$^{-1}$, where the subscript $S$ and $L$ indicate smaller and larger values of the scale radius and velocity. In the inner part of a galaxy, the mass is more sensitive to the scale radius with $R_S$ leading to a larger mass, while at large radii the mass is more sensitive to the scale velocity with $V_L$ leading to a larger mass. 

Our GC data covers a large region of M87 and so we have obtained a robust solution for the free parameters. 
However we still have some degeneracy diagonally. The two halos with $R_{S}, V_{S}$ and $R_{L}, V_{L}$ have similar mass distributions. The former has a larger mass at $R \sim 40$ kpc and a smaller mass at higher radii, and conversely so for the latter. The ill-understood large velocity dispersion at $R\sim 40$ kpc may be part of the reason for this degeneracy (see Peng et al. 2014 for more details). 

We can see that the models with parameters diagonally distributed have even smaller difference on the mass distribution in the inner part. Data only extending to $R< 50$ kpc, as in previous investigations, have a very weak ability to distinguish these kinds of models, and that may be part of the reason that M2011 and G2009 obtained very different dark matter distributions even when they used the same GC data.

In our M87 investigation, we have focused on a logarithmic potential for the dark matter distribution.  It must be noted however that the widely used NFW profile suffers similar degeneracy issues, as discussed in the appendix. Dark matter potentials are usually assumed to be spherical. If we allow a variation of the dark matter halo shape, the shape will also be degenerate radially.  

Much effort has been devoted to constraining the dark matter distribution of elliptical galaxies. IFU data extending to $\sim 2R_e$ like VIRUS-p is helpful to remove mass-anisotropy degeneracy near the centre, but has limited constraining power on the dark matter distribution. 
As shown in Fig.~\ref{fig:mass-M87}, the discrepancy between M2011 mass and our mass is less than $20\%$ at $r< 15$ kpc, but increases to be $60\%$ at $r= 40$ kpc. If we extrapolate the M2011 mass out to $r=180$ kpc, the discrepancy increases to be $100\%$.
As we have illustrated above, model parameters in the mass distribution are degenerate and extrapolation of the mass profile to larger radii is not reliable. Discrete data extending to much larger radii than IFU data normally extend (e.g. SAURON \citealt{Bacon2001}, CALIFA \citealt{CALIFA2012}) are critical to constraining the dark matter distributions for most galaxies. 

\subsection{Uncertainties in our model parameters}

Uncertainties in our model parameters arise from three main sources: 1) the stellar surface brightness profile used is not as extended spatially as the new NGVS profile, 2) the GC surface density profile is assumed to be spherical when it actually appears to be elliptical, 3) a spherical dark matter halo is assumed. We  discuss these three sources in turn.

The new NGVS surface brightness profile differs by $5\% - 20\%$ from the Cappellari et al. (2006) profile we used at $40<R<80$ kpc, where the luminous matter fraction is $30\%-15\%$. Using the NGVS surface brightness would have caused a difference of $\sim 3\%$ in the total mass at $40<R<80$ kpc, with even smaller differences in other regions. The effect on the mass-to-light ratio would be very small.

\begin{figure}
\centering\includegraphics[width=\hsize]{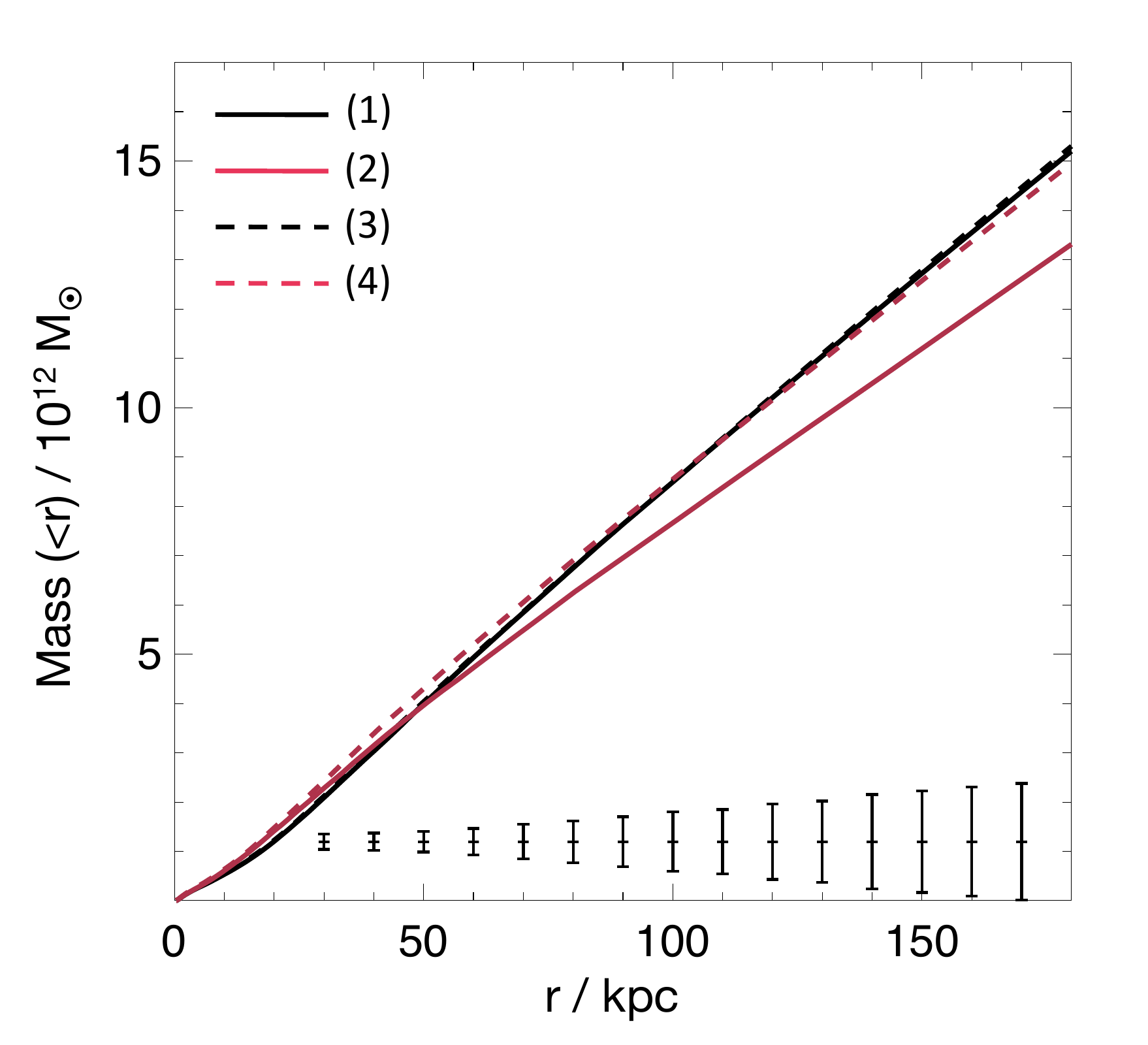}
\caption{Mass profiles for the four Jeans models. For the elliptical halos, ``radius'' is taken to be $\sqrt ab$ where a,b are the major and minor axis values. The pluses represent the typical error bars at different radii for the mass profiles. The maximum mass differences between the models are $\approx 10\%$.
}
\label{fig:jam}
\end{figure}

The recent NGVS result (Durrell et al. 2014, submitted) shows that the GC distribution is not spherical, as shown in Fig.~\ref{fig:SB}, but is round in the centre, becoming more elliptical at large radii. The GC number density does not affect the potential, as the mass of the GCs is negligible compared to that of the stars in the main galaxy. The initial spatial distribution of the M2M test particles in the GC models will be affected, as will the number of particles used to construct model observables. We do not expect the GC distribution to affect the results significantly (see Fig.~\ref{fig:jam}).

Recent cosmological simulations show that the shape of a dark matter halo may be correlated with the shape of the luminous matter distribution it contains, with the dark halo usually being rounder than the luminous matter \citep{Wu2014}. The stellar distribution and GC distribution in Fig.~\ref{fig:SB} show that M87 is quite round in the centre, with the axial ratio $q$ becoming smaller at large radii. The flattening of the logarithmic dark matter potential may be very roughly estimated as $2/3+q_{dm}/3$ (\citealt{BT2008}, p. 77), where $q_{dm}$ is the axial ratio of the dark matter halo.  As a consequence we do not expect the flattening of the potential to be that severe. 

In order to assess quantitatively the effects of points 2) and 3), we run isotropic axisymmetric Jeans models utilising the discrete GC data, following the approach in \cite{Watkins2013}.  To be consistent with the M2M models, the potential is a combination of a luminous matter distribution and a dark matter halo. The luminous matter distribution is constructed from the (deprojected) surface brightness profile multiplied by a constant mass-to-light ratio. We take the ratio as $M/L_I = 6.0$ as obtained from the M87 M2M models. The dark matter halo is a logarithmic halo with two parameters, the scale velocity $V_s$ and the scale radius $R_s$ as before, and a third optional free parameter, the axial ratio $q_{dm}$ . The GC number density profile is required for the calculation of the velocity moments.  We create four sets of Jeans models with the following conditions, 
\begin{enumerate}
\item a spherical GC number density and a spherical dark matter potential, 
\item an elliptical GC number density with known flattening and a spherical dark matter potential, 
\item a spherical GC number density and an axisymmetric dark matter potential with the flattening $q_{dm}$ as a free parameter, 
\item an elliptical GC number density and an axisymmetric dark matter potential with $q_{dm}$ again free.
\end{enumerate}

The mass profile of the best models are shown in Fig.~\ref{fig:jam}, these four sets of Jeans models only have minor differences in the total mass of the system. In particular, the models with an elliptical GC surface number density have a smaller mass with up to a $10\%$ difference.  The models favour a round dark matter halo with $q_{dm}$ approaching unity. The flattening of a dark matter halo is correlated with scale velocity, but its effect on the total mass is small. We
expect that were the analysis to be performed using M2M modelling the results would not be dissimilar. A more rigorous treatment will be the subject of a future investigation.

\section{Conclusion}
We have created M2M models to study the dynamics of
the giant elliptical galaxy M87. 
We model with discrete GC data and SAURON IFU data in
separate sets of models. 
As part of the modelling we have used a smooth logarithmic dark matter halo, 
while the M87 luminous matter distribution (excluding GCs) is constructed 
from a multi-Gaussian Expansion fit, assuming a constant stellar mass-to-light ratio. 
The surface number density of GCs (fitted by a Sersic profile)
and the stellar surface brightness are used as tracer densities
of the GC and SAURON models respectively. 
GC velocities are used directly in the model
as discrete data and are not binned prior to modelling.
With the 263 new GC velocities from NGVS survey, we
have modelled with 896 GC discrete LOS velocity measurements extending to $R \sim 180$ kpc. For the M87 GC
models, we use regularization to reduce fluctuations in
the model log likelihood functions.

The mass of M87 is found to be $1.5\pm0.2\times10^{13} M_{\odot}$ within 180 kpc. Degeneracies within the dynamical models have been partly removed since we have GC constraints at large radii. Our three key parameters are well determined with the dark matter scale velocity $V_s = 591 \pm 50$ km s$^{-1}$, the dark matter scale radius $R_s=42 \pm 10$ kpc, and the stellar mass-to-light ratio $M/L_{\mathrm I} = 6.0\pm0.3 $ $M_{\odot}/L_{\odot}$.

Previous results derived from M87 GCs (mostly with $R<50$ kpc) (e.g. \citealt{Murphy2011}; \citealt{Gebhardt2009}) give estimates for M87's mass generally larger than our value.
These earlier, larger mass values may be the result of the large velocity dispersion at $R\sim 40$ kpc in the GC data. This large dispersion may be caused by kinematic substructures. 
On the other hand, the only previous results obtained by GCs at large radius (extending to 150 kpc but with less data point than our sample), the mass obtained by \cite{Strader2011} by Jeans analysis is smaller than our mass at $R \sim 85$ kpc.

 The mass we derive is in good agreement with that inferred from \emph{ROSAT} X-ray observation out to 180 kpc. Within 30 kpc our mass is also consistent with that inferred from \emph{Chandra} and \emph{XMM-Newton} X-ray observations, while within 120 kpc it is about $20\%$ smaller. 

The velocity dispersion anisotropy of the best-fitting GC model is small and is consistent with the M87 stellar anisotropy in the regions of overlap. This suggests that M87 may have been formed by multiple mergers which made the motions isotropic. Our result places strong constraints on the orbital distribution of GCs.

Different kinds of data play different roles within the kinematical models. The stellar mass-to-light ratio is mainly constrained by the SAURON IFU data, while the GC velocities at large radii are critical to constrain the dark matter distribution. 
In previous models, the lack of GC data at $R>50$ kpc affected the results in two ways. Firstly, the large velocity dispersion at $R\sim40$ kpc probably causes an overestimate of the total mass. Secondly, the degeneracy between the dark matter parameters ($V_s$ and $R_s$) is difficult to remove, which is part of the reason why M2011 and G2009 obtained very different dark matter distributions (see Fig~\ref{fig:mass-M87}). With GCs extending to $R=180$ kpc, we can better overcome these problems and remove the degeneracy between parameters in the dark matter distribution, and thus partly remove the degeneracy between the dark matter distribution and stellar mass-to-light ratio.  

The current investigation can be improved in several respects. Firstly, since the work was completed, more M87 GC velocities have become available due to the ongoing NGVS programme. Secondly, GCs in M87 can be divided into blue and red groups \citep{Cote2001} which behave somewhat differently in their kinematics \citep{Strader2011}.  It would be interesting to investigate their dynamics separately with the M2M method. Thirdly, the stellar surface brightness distribution of M87 is more extensively sampled by recent ACS+NGVS data (Ferrarese et al., in preparation) and would improve modelling of the stellar component. Fourthly, the flattening of the GC distribution and the dark matter halo can be accommodated in our modelling.  Lastly, the effect of the central black hole can be taken into account in a straightforward manner. Whilst preliminary studies indicate that none of these will change our results significantly, we will address these matters in a future investigation to provide even more precise constraints on the dynamical structure of M87.

\section*{Acknowledgments}
The authors thank Ortwin Gerhard and Glenn van de Ven for useful discussions.
Computer runs were mainly performed on the Laohu high performance computer cluster and BH cluster of the National Astronomical Observatories, Chinese Academy of Sciences (NAOC). We thank Profs. Rainer Spurzem and Youjun Lu for enabling our access to these clusters. SM and RJL acknowledge the financial support of the Chinese Academy of Sciences and NAOC. This work has also been supported by the Strategic Priority Research Program ``The Emergence of Cosmological Structures'' of the Chinese Academy of Sciences Grant No. XDB09010500 (EWP and SM), and by the National Natural Science Foundation of China (NSFC) under grant numbers 11333003 (SM and RJL), 11173003 (EWP and BL) and 11203017 (CL). 

The NGVS team thanks the directors and the staff of the Canada-France-Hawaii Telescope: their commitment and ingenuity have helped the survey become a reality. This work has been supported in part by the French Agence Nationale de la Recherche (ANR) Grant Programme Blanc VIRAGE (ANR10-BLANC-0506-01), and by the Canadian Advanced Network for Astronomical Research (CANFAR) through funding from CANARIE under the Network-Enabled Platforms program. This research has made use of the facilities at the Canadian Astronomy Data Centre, which are operated by the National Research Council of Canada with support from the Canadian Space Agency.

\bibliography{ms_apj} 

\clearpage

\appendix

\section{Testing the M2M modelling procedure with NGC 4374}
\label{app:4374}
As well as GCs, PN discrete data are a good tracer of the gravitational potential of elliptical galaxies. Within the galaxies with publicly available PN velocity data, NGC 4374 has the most ($\sim450$) PNs with good velocity measurements \citep{Coccato2009}. There is a Jeans analysis for these data which implies a standard NFW halo for the galaxy \citep{Napolitano2011}. We consider the NGC 4374 PN data to be a good data set to test our procedure for modelling discrete data, and aim to provide a cross check on the results from \cite{Napolitano2011}.

\subsection{The spherical approach}
Most elliptical galaxies have a surface brightness profile which can be well fitted by a Sersic profile (\citealt{Sersic1968}; \citealt{Capaccioli1989}; \citealt{Caon1993}; \citealt{Graham2003}; \citealt{Ferrarese2006}; \citealt{Cote2007}). 
We use the Sersic profile (see Eq.~\ref{eqn:sersic}) as a starting point for creating a spherical potential from the inverse Abell transform of the 1D surface brightness profile \citep{BT2008}. 

The luminosity density $j(r)$ is derived from the inverse Abell transform
\begin{equation}
j(r) = -\frac{1}{\pi} \int_r^{\infty}\frac{dI}{dR} \frac{dR}{\sqrt{R^2 - r^2}},
\end{equation}
where $I$ is the fitted Sersic surface brightness profile.
The mass
\begin{equation}
M(r) = 4\pi (M/L) \int_0^r j(s) s^2 ds,
\end{equation}
and potential
\begin{equation}
\Phi(r) = -4 \pi G (M/L) \Bigr[ \frac{1}{r}\int_0^r j(s)s^2 ds + \int_r^{\infty}j(s)sds  \Bigr],
\end{equation}
are obtained directly from $j(r)$. The stellar mass-to-light ratio is $M/L$ and $G$ is the gravitational constant.

Multiple integrations are needed to obtain the potential from the surface brightness profile so, for performance reasons, we use interpolation tables during modelling. The potential and its associated accelerations are pre-calculated and stored in tables. Linear interpolation is used on the tables to obtain the required values.

The Sersic profile of the main body of a galaxy may not fit the very centre of the galaxy well due to the presence of a `core' or a `cusp' at the centre. This effect usually extends to a few arc-seconds \citep{Ferrarese2006}, and affects the calculation of the potential. If the kinematic data in use reaches the region within the influence radius of the `core' or `cusp', we add a linear function in the centre to approximate the real surface brightness profile. 

\subsection{NFW dark matter halo}
The NFW profile \citep{Navarro1996} is a prediction from cold dark matter models. The profile has two free parameters, the virial mass $M_v$ and concentration parameter $C$. The virial mass is the mass within the virial radius $r_v$. The average density is $c_v  \rho_c^0$, where $\rho_c^0 = 1.37 \times 10^{-7} {\mathrm M}_{\odot} {\mathrm pc}^{-3}$ is the critical density, and $c_v$ takes the value of 178 \citep{Lokas&Mamon2001}. (We take the Hubble constant as $H_0=70$km/s/Mpc.) 

The virial mass is
\begin{equation}
 M_v = 4/3 \pi r_v^3 c_v \rho_c^0,
\end{equation}
and the mass within radius $r$ is
\begin{equation}
M(<x) = M_v g(C) \Bigr[ \ln(1+C x) - \frac{C x}{1 + C x}   \Bigr],
\end{equation}
where $x = \frac{r}{r_v}$, and $g(C) = \frac{1}{\ln(1+C) - C/(1+C)}$.

The gravitational potential is
\begin{equation}
\Phi(x) = - V_v^2 g(C) \frac{\ln(1 + C x)}{x},
\end{equation}
where $V_v^2 \equiv V(r_v)^2 = \frac{G M_v}{r_v} $.

$M_v$ and $C$ are not independent. In the collisionless cold dark matter model, there is a relationship found from numerical simulations \citep{Maccio2008}
\begin{equation}
\label{eqn:MvC}
C(M_v) \approx 12 \Bigr(\frac{M_v}{10^{11} M_{\odot}}\Bigr)^{-0.094},
\end{equation}
where the $1\sigma$ dispersion is 0.11 dex.

In M2M models using the NFW dark matter profile, we have three, free parameters, the stellar mass-to-light ratio $M/L$, the virial mass $M_v$, and the concentration parameter $C$.

\subsection{M2M Modelling}
We create spherical M2M models for NGC 4374 to test how the discrete data constraint works in a M2M model.
The surface brightness profile in the $V$-band to $6.7 R_e$, the long slit velocity dispersion profile within $1R_e$, and 450 PNs out to $6R_e$ are used as model constraints. We use $R_e = 6.02$ kpc and $L_V = 7.64\times10^{10} L_{V,\odot}$ following \cite{Napolitano2011}.
The normalized PNs number density profile is consistent with the stellar surface brightness. 
The velocity dispersion of the PNs is about $20\%$ smaller than the stellar velocity dispersion in the overlap region. This may be due to the limited number of PN velocities at this region. We assume that the PNs and stars follow the same distribution function and model them together. 
The data are generally the same as \cite{Napolitano2011} used for their Jeans analysis for the galaxy. The only difference is that we include all the PNs in our model as constraints, while \cite{Napolitano2011} has binned the data and simply excluded the PN data within the overlap regions.

The initial M2M models are set up with $N=500000$ particles extending to $20R_e$, with particle initial conditions set following \S~\ref{sec:ini}.

The gravitational potential is an NFW dark matter halo plus the luminous matter potential. As for M87, we ignore the effects of any central black hole.
The luminous matter potential is constructed from the fitted Sersic surface brightness profile ($R_0 = 113.5'', n = 6.11, b_n = 5.16$) following \S\ref{sec:MGE}. 

The M2M model has three, free parameters: the stellar mass-to-light ratio $M/L_{\mathrm V}$, the virial mass $M_v$, and the concentration parameter $C$. We run a series of models with a parameter grid of $12\times5\times6$ on $M_v \times C \times M/L_{\mathrm V}$. 

\begin{figure}
\centering\includegraphics[width=8cm,height=7.2cm]{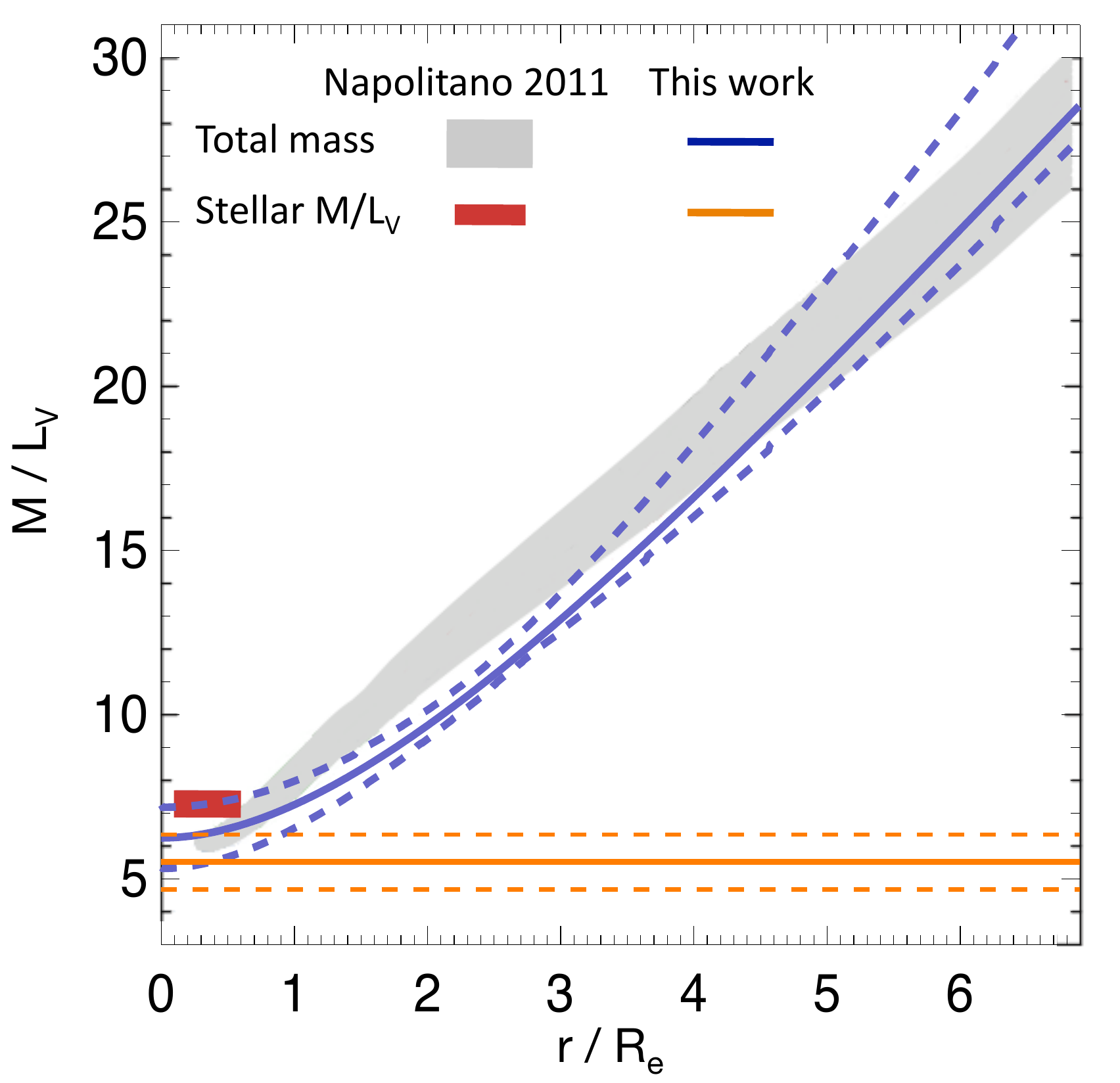}
\caption{The total mass-to-light ratio (in units of $M_{\odot}/L_{V,\odot}$) profile of NGC 4374. The dark blue solid and dashed lines are the total mass-to-light ratio profile of the minimum 
$\Delta G$ model and its uncertainty. The orange solid and dashed lines are the stellar $M/L_{\mathrm V}$ and uncertainty. The grey and red dashed regions are the corresponding total and stellar mass-to-light ratios of \protect\cite{Napolitano2011}. }
\label{fig:massNGC4374}
\end{figure}

Given the constraints are surface brightness, the long slit velocity dispersion and the PN discrete velocity data,
we take $G(\mathbf{p})$ (see equation~\ref{eqn:Gdef}) as
\begin{equation}
G(\mathbf{p}) = -\frac{1}{2} \left( \lambda_{\mathrm {SB}}\chi^2_{\mathrm {SB}} + \lambda_{\mathrm {VD}} \chi^2_{\mathrm {VD}} \right) + \lambda_D \mathcal{L},
\end{equation}
where the $\mathrm {SB}$ and $\mathrm {VD}$ terms are self-explanatory and $\mathcal{L}$ is the log likelihood function for the PN discrete velocity data.  The values of the $\lambda$ parameters are $\lambda_{\mathrm {SB}} = 10^{-3}$, $\lambda_{\mathrm {VD}} = 2\times 10^{-3}$ and $\lambda_D = 4 \times 10^{-3}$.

Using our M2M models, we constrain the mass of NGC 4374 to within a $\sim 25\%$ uncertainty at $r<6R_e$. However, we can not obtain tight constraints on the model parameters $C$, $M_v$ and $M/L_V$. The dark matter and luminous matter are degenerate, and, for the dark matter, the mass distribution of an NFW halo is more sensitive to $C$ in the region with dynamical data and the influence of $M_v$ is weak. If we want to obtain better constraints on the model parameters for NGC 4374, the dynamical data must extend to larger radii where the mass distribution is more sensitive to $M_v$, then we can partly remove the degeneracy.

Fig.~\ref{fig:massNGC4374} shows the $1\sigma$ error of the mass distribution.  We have constrained the total mass of NGC 4374 to be $1.89^{+0.27}_{-0.08} \times 10^{12} M_{\odot}$ with $6R_e$. The $1 \sigma$ error bar is derived from models within the confidence level $\Delta G = 0.006$, where the criterion $0.006$ is the $\Delta G$ fluctuation round the minimum $\Delta G$ (see \S~\ref{sec:paramest}).   
These models within $1\sigma$ error roughly have the concentration parameter $C$ in the range $7-16$, virial mass $M_v$ in the range $(100-350)\times 7.6\times 10^{10} M_{\odot}$, and stellar mass to light ratio $M/L_{\mathrm V}$ in the range $4-8$. 
The mass at $R< 6R_e$ is generally consistent with that obtained from the Jeans analysis of \cite{Napolitano2011}. The stellar mass-to-light ratio we obtain is slightly smaller, and may be affected by the PNs inside $1R_e$ which were simply excluded in \cite{Napolitano2011}. Thus, the total mass inside $2R_e$ we obtain is also slightly smaller. 

\begin{figure}
\centering\includegraphics[width=8cm,height=4.7cm]{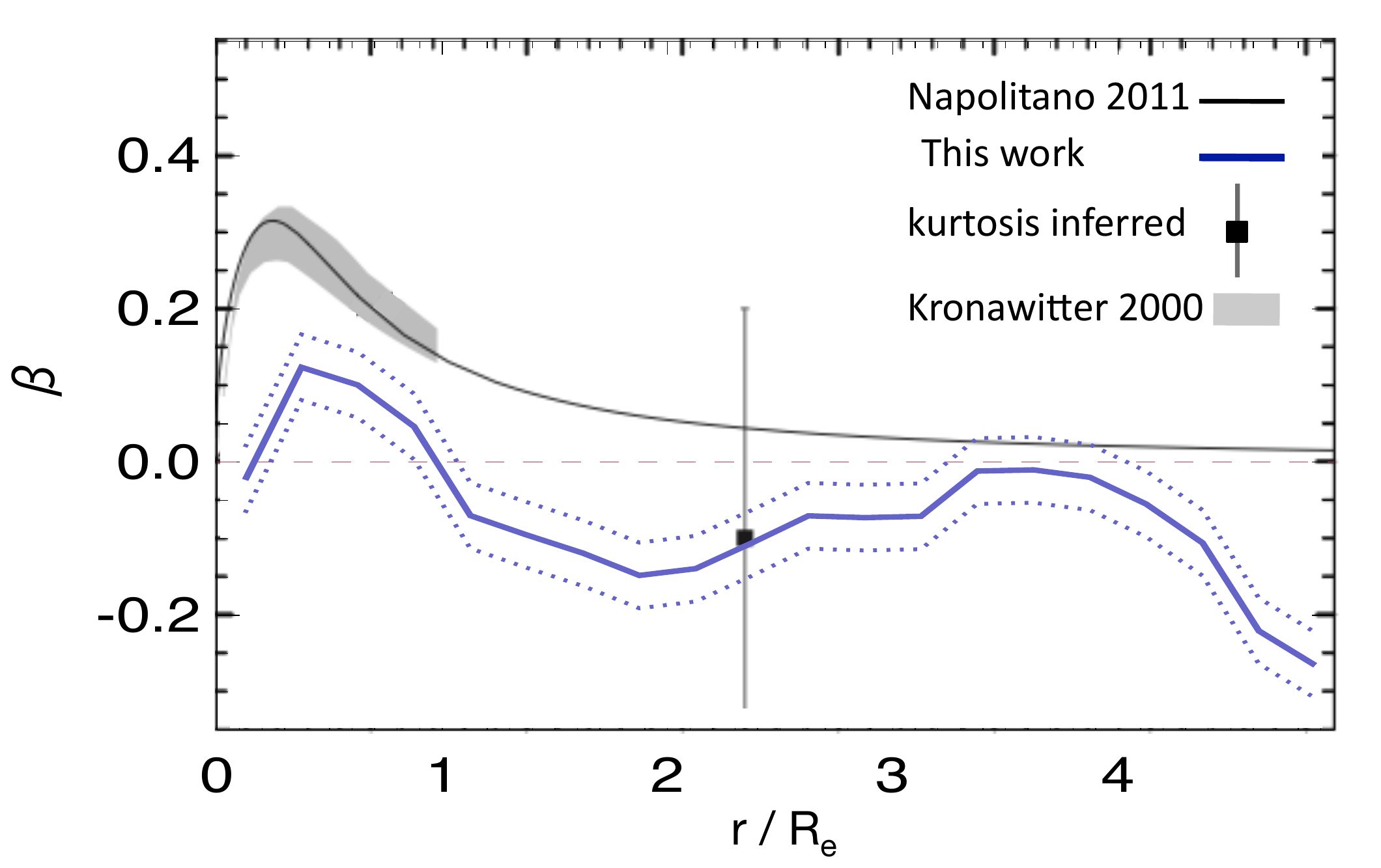}
\caption{The $\beta(r)$ profile. The dark blue solid and dashed lines are the $\beta$ profile and uncertainties of the minimum $\Delta G$ model. The heuristic $\beta(r)$ of NGC 4374 from a Jean analysis is shown in black solid line \protect\citep{Napolitano2011}, and the grey dashes are the modelled $\beta$ profile from \protect\cite{Kronawitter2000}. The filled square with $1\sigma$ error bars is the anisotropy value derived from direct kurtosis inferences \protect\citep{Napolitano2011}.}
\label{fig:betaNGC4374}
\end{figure}

Fig.~\ref{fig:betaNGC4374} shows that the velocity dispersion anisotropy $\beta$ profile of the best model has a shape similar to \cite{Napolitano2011}.
However, our $\beta$ is $\sim 0.1$ is systematically smaller, with a slightly smaller mass. What we have found is a degenerate solution of \cite{Napolitano2011}. This may be another effect contributing towards a smaller stellar mass-to-light ratio.

If the system is assumed to have a constant dispersion profile as we have here for NGC 4374, the LOS velocity distribution kurtosis $\kappa$ is then a simple matter of projection effects for a given $\beta$ and luminosity profile. Thus its internal anisotropy $\beta$ can be directly obtained without any need for dynamical modelling \citep{Napolitano2009}. The black square with error bar in Fig.~\ref{fig:betaNGC4374} indicates the $\beta$ value of NGC 4374 obtained in this way \citep{Napolitano2011}, and this matches our $\beta$ value although the error bar is large.
In the Jeans analyses, \cite{Napolitano2011} assumed an analytic $\beta$ profile which prevents the model having $\beta < 0$. In our models, we did not make any assumptions about the anisotropy, the model naturally producing the anisotropy profile as constrained by the discrete data. 

We have shown above the success of our method in modelling the discrete velocity data for PNs in NGC 4374. The M2M results are understandable and acceptable in the context of the physical scenario being modelled. We are confident that our procedure can be applied to the M87 globular cluster system.

\end{document}